\documentclass{aa}  

\usepackage{graphicx}
\usepackage{txfonts}
\usepackage{orcidlink}
\usepackage{booktabs}
\usepackage{upgreek}
\usepackage{threeparttable}
\usepackage[T1]{fontenc}
\usepackage{comment}
\DeclareRobustCommand{\VAN}[3]{#2}
\let\VANthebibliography\thebibliography
\def\thebibliography{\DeclareRobustCommand{\VAN}[3]{##3}\VANthebibliography}
\usepackage{graphicx}	
\usepackage{amsmath}	
\usepackage{tablefootnote}
\usepackage{url}
\bibpunct{(}{)}{;}{a}{}{,} 
\usepackage{txfonts}

\usepackage{hyperref}
\hypersetup{
    colorlinks=true,
    linkcolor=blue,
    citecolor=blue,
    filecolor=blue
}

\begin{document}

   \title{Exploring magnetised galactic outflows in starburst dwarf galaxies NGC\,3125 and IC\,4662}
    \titlerunning{Magnetised Galactic Outflows in Dwarfs}

   \author{Sam Taziaux \inst{\ref{rub},\ref{rapp}}\orcidlink{0009-0001-6908-2433}
          \and
          Ancla Müller \inst{\ref{rub}}\orcidlink{0000-0001-9184-7845}
          \and
          Björn Adebahr \inst{\ref{rub}}\orcidlink{0000-0002-5447-6878}
          \and
          Aritra Basu \inst{\ref{TLS}, \ref{mpi_bonn}}\orcidlink{0000-0003-2030-3394}
          \and
          Christoph Pfrommer \inst{\ref{pots}}\orcidlink{0000-0002-7275-3998}
          \and
          Michael Stein \inst{\ref{rub}}\orcidlink{0000-0001-8428-7085}
          \and
          Krysztof T. Chyży \inst{\ref{pol}}\orcidlink{0000-0002-6280-2872}
          \and
          Dominik J. Bomans \inst{\ref{rub}, \ref{rapp}}\orcidlink{0000-0001-5126-5365}
          \and 
           Torsten Enßlin \inst{\ref{mpi_garch}}\orcidlink{0000-0001-5246-1624}
          \and
          Volker Heesen \inst{\ref{HH}}\orcidlink{0000-0002-2082-407X}
          \and
          Peter Kamphuis \inst{\ref{rub}}\orcidlink{0000-0002-5425-6074}
          \and
          Marian Soida \inst{\ref{pol}}\orcidlink{0000-0002-1671-9285}
          \and
          Marek Wezgowiec \inst{\ref{pol}}\orcidlink{0000-0002-4112-9607}
          \and
          Ralf-Jürgen Dettmar \inst{\ref{rub}, \ref{rapp}}\orcidlink{0000-0001-8206-5956}
          \and
          Samata Das \inst{\ref{tp4}}\orcidlink{0009-0005-9140-7811}
          \and
          Julia Tjus\inst{\ref{rapp}, \ref{tp4}, \ref{sweden}}\orcidlink{0000-0002-1748-7367}
          }

   \institute{Ruhr University Bochum, Faculty of Physics and Astronomy, Astronomical Institute (AIRUB), Universitätsstraße 150, 44801 Bochum, Germany
              \email{sam.taziaux@rub.de} \label{rub}
        \and 
             Ruhr Astroparticle and Plasma Physics Center (RAPP Center) \label{rapp}
        \and 
             Th\"uringer Landessternwarte, Sternwarte 5, 07778 Tautenburg, Germany \label{TLS}
         \and
             Max-Planck-Institut für Radioastronomie, Auf dem H\"ugel 69, 53121 Bonn, Germany \label{mpi_bonn}
        \and
            Leibniz-Institute for Astrophysics Potsdam (AIP), An der Sternwarte 16, 14482 Potsdam, Germany \label{pots}
        \and
            Astronomical Observatory of the Jagiellonian University, ul. Orla 171, 30-244 Kraków, Poland \label{pol}
        \and
            Max-Planck-Institut für Astrophysik, Karl-Schwarzschild-Str. 1, 85748 Garching, Germany \label{mpi_garch}
        \and
            Hamburger Sternwarte, Universität Hamburg, Gojenbergsweg 112, 21029 Hamburg, Germany \label{HH}
        \and
            Ruhr University Bochum, Faculty for Physics \& Astronomy, Theoretical Physics IV: Plasma Astroparticle Physics, 44780 Bochum, Germany \label{tp4}
        \and 
            Department of Space, Earth and Environment, Chalmers University of Technology, Gothenburg, Sweden \label{sweden}
            }

   \date{}

  \abstract
   {The study of radio emission in starburst dwarf galaxies provides a unique opportunity to investigate the mechanisms responsible for the amplification and transport of magnetic fields. Local dwarfs are often considered proxies for early-Universe galaxies, so this study may provide insights into the role of non-thermal components in the formation and evolution of larger galaxies.}
   {By investigating the radio continuum spectra and maps of the starburst dwarf galaxies, we aim to draw conclusions on their magnetic field strengths and configurations, as well as the dynamics of cosmic ray (CR) transport.}
   {We perform a radio continuum polarimetry study of two of the brightest starburst IRAS Revised Bright Galaxy Sample (RBGS) dwarf galaxies, NGC\,3125 and IC\,4662. By combining data of the Australian Telescope Compact Array (2.1\,GHz) and MeerKAT (1.28\,GHz), we analyse the underlying emission mechanism and the CR transport in these systems.}
   {We find flat spectra in those dwarf galaxies over the entire investigated frequency range, which sharply contrasts with observations of massive spiral galaxies. Because the expected cooling time of CR electrons is much shorter than their escape time, we would expect a steepened steady-state CR electron spectrum. The flat observed spectra suggest a substantial contribution from free-free emission at high frequencies and absorption at low frequencies, which may solve this puzzle. For NGC\,3125, we measure a degree of polarisation between $0.75\,\%$ and $2.6\,\%$, implying a turbulent field and supporting the picture of a comparably large thermal emission component that could be sourced by stellar radiation feedback and supernovae. }
   {}


   \maketitle

\section{Introduction}
Dwarf starburst galaxies serve as important laboratories for analysing the effects of feedback processes, such as star formation, on galactic evolution. Their low escape velocities mean that even a modest starburst event can lead to winds \citep[e.g.,][]{chyzy_magnetized_2016} or outflows \citep[e.g.,][]{Adebahr_2013}. 
Understanding the complexities of starburst processes in dwarf galaxies is therefore an important step in understanding the feedback period processes between star-formation and galactic evolution \citep[e.g.,][]{Dekel_1986, Stevens_2002, chyzy_magnetized_2016}. 

Two challenging questions in magnetic field research are understanding the role of magnetic fields in dwarf galaxies and how these fields formed and evolved in the early Universe \citep[e.g.,][]{Pakmor_2017, Pakmor_2024, Liu_2022}. 
Radio continuum observations provide insights into magnetic field strength \citep{beck_2019} with limitation of starburst galaxies \citep[e.g.,][]{Lacki_2013,Ruszkowski_2023}, field orientation through polarisation and Rotation Measure (RM) synthesis \citep{Brentjens_2005,Pakmor_2018,Reissl_2023}, and the relation between ordered and total magnetic fields in galaxy disks and halos \citep{Beck_2005}. Furthermore, they enable the investigation of cosmic ray (CR) transport, using relativistic electrons as tracers \citep[e.g.,][]{Lacki_2010,Werhahn_2021,Pfrommer_2022,heesen_nearby_2022}. CRs propagate from their origin in supernova remnants through the galactic disk into the halo via advection, diffusion and streaming mechanisms \citep[e.g.,][]{Strong_2007,Stein_2019,Thomas_2020,Stein_2023}. 
CRs excite Alfv\'en and whistler waves through resonant plasma instabilities, which in turn scatter CRs, thereby lowering their effective drift speed along the mean magnetic field \citep{Kulsrud_1969,Shalaby_2021,Shalaby_2023,Lemmerz_2024}. Hence, CRs primarily interact with the galactic magnetic field, transferring momentum to the thermal gas through resonant scattering at plasma waves. As a result, CRs apply a pressure on the ambient plasma, which is the basis of formulations of CR hydrodynamics \citep{Zweibel_2013,Pfrommer_2017,Thomas_2019}. Galactic wind models incorporating this hydrodynamic description indicate that CR transport significantly influences mass-loss rates, gas distribution, and wind formation \citep[e.g.,][]{Breitschwerdt_1991,Uhlig_2012,Pakmor_2016,Girichidis_2016,Girichidis_2018, Recchia_2017, Dashyan_2020, Thomas_2023}. 
\citet{Thomas_2024} model CR transport in a multiphase interstellar medium in a global Milky Way-like galaxy and suggest that CRs have a significant impact on the formation of galactic winds due to their long cooling times and better plasma coupling compared to radiation. This makes galactic winds denser and increases the mass loading factor in these CR-driven winds, thereby providing more effective feedback on galaxies by moving this gas to larger heights in the halo and reducing the amount of gas available for future star formation.
Outflow describes any material moving away from a central object, whereas wind specifically refers to gas ejected from a galaxy, typically driven by mechanisms such as radiation pressure, and can encompass both diffuse flows and jets \citep{Veilleux_2005}.

We focus on further exploring the properties of these winds, such as wind speed and magnetic field strength in the galactic halo, driven by stellar feedback. Because of the very low radiative loss rates of CR protons (CRPs), they only can be observed via the decay products in hadronic collisions with the ambient gas, which scales with the gas density and thus mostly illuminates the dense gaseous phase \citep{Pfrommer_2004,Pfrommer_gamma_2017,Werhahn_2023}. To study the lower density outflow regions, we have to turn to the radiative products of CR electrons (CREs), in particular the radio synchrotron emission. In this context, CREs need to be modelled alongside CRPs, which collectively form the entire CR population. The different loss timescales of (relativistic) electrons and protons cause a deviation of their energy spectra \citep{Ruszkowski_2023} and hence, it is critical to model their spectra separately for providing a realistic prediction of the radio emission in galaxies \citep{Chiu_2024}. 

In recent years, various studies have been conducted to better understand the radio spectrum's shape in dwarf galaxies \citep[e.g.,][Taziaux et al., in prep]{klein_2018}. 
The few in-depth radio continuum and polarisation studies of dwarf galaxies so far have shown --- (1) a mean spectral index ($S\propto \nu^{\alpha_\text{nth}}$) of $-0.6$ that steepens at the galaxy outskirts to $-1.1$ \citep[e.g.,][]{chyzy_regular_2000,kepley_role_2010,Kepley_2011,chyzy_magnetized_2016,westcott_spatially_2018}, (2) a mean equipartition field of 13.5\,$\upmu$G to 38\,$\upmu$G \citep[e.g.,][]{kepley_role_2010,Kepley_2011,chyzy_magnetized_2016, Basu_2017}, (3) they can host a large-scale magnetised halo up to a few kiloparsecs \citep[e.g.,][]{chyzy_magnetized_2016,chyzy_regular_2000} being partly polarised reaching an ordered field strength of about $6\textrm{--}8\,\upmu$G and a degree of polarisation of $50\,\%$, providing important information about the structure and strength of the magnetic field \citep{chyzy_magnetized_2016}, and (4) that within the resolved star-forming knots the spectral index can reach a flatter spectrum of $-0.3$ and a significantly lower degree of polarisation of about $3\,\%$ due to a combination of Faraday depolarisation caused by the high densities of ionised gas and magnetic fields in these regions \citep[e.g.,][]{chyzy_magnetized_2016,kepley_role_2010,chyzy_regular_2000, Basu_2017}. 
Taking the recent findings of the magnetic field properties of dwarf galaxies into account, they are considered the best candidates for cosmic magnetisers in the early Universe \citep[e.g.,][]{Atek_2024, Sanati_2024}, as they resemble the primordial galaxies during the Epoch of Reionisation \citep{Sanati_2024}. The magnetisation of the intergalactic medium via galactic outflows can be important as it can affect the formation and evolution of galaxies and help us to understand the structure and evolution of the Universe \citep[e.g.,][]{chyzy_magnetized_2016,kepley_role_2010,Pakmor_2020,chyzy_regular_2000,Cannon_2005,klein_1996}.

In this paper, we present a radio continuum polarimetry study of the two brightest IRAS Revised Bright Galaxy Sample (RBGS) dwarf galaxies, NGC\,3125 and IC\,4662, that allows us to determine the magnetic field strengths and the structure in these galaxies. 
These datasets represent the most sensitive set of radio continuum observations of NGC\,3125 and IC\,4662 to date, and the first observations of their polarised emission at 16\,cm. We describe our data set in Sect.~\ref{data} and present the basic properties of radio continuum of both dwarf galaxies in Sect.~\ref{radiocont}. We place our observations in the context of what is known of NGC\,3125 and IC\,4662 in Sect.~\ref{discussion}. Finally, in Sect.~\ref{conclusion}, we present a summary of our work and our conclusions. In Appendices~\ref{NT}, ~\ref{tf_sect} and \ref{spix_nth}, we provide the non-thermal, the thermal fraction and the non-thermal spectral index maps of NGC\,3125 and IC\,4662, respectively. In Appendices~\ref{MF} and \ref{frm}, we investigate the properties of the magnetic field of NGC\,3125 and IC\,4662. In Appendix~\ref{timescale}, we estimate the different timescale for CRE losses. Throughout the paper, we use the Gaussian cgs unit system.

\section{Observation and data reduction}
\label{data}

\begin{table*}
\caption{Basic properties of NGC\,3125 and IC\,4662.}
\label{Tab: properties}
\centering
\begin{threeparttable}
\begin{tabular}{lccc}
\toprule
& NGC\,3125 & IC\,4662 & Reference\\
\midrule
RA (J2000) & 10$^{\rm h}$\,06$^{\rm m}$\,33$^{\rm s}$.372 & 17$^{\rm h}$\,47$^{\rm m}$\,08$^{\rm s}$.86 & LEDA database\\
Dec (J2000) & -29$^{\rm d}$\,56$^{\rm m}$\,05$^{\rm s}$.50 & -64$^{\rm d}$\,38$^{\rm m}$\,30$^{\rm s}$.3  & LEDA database \\
Type & BCD (WR) & IBm (WR) & \citet{Hadfield_2006,Crowther_2009}\\
Inclination / $^{\circ}$ & 55.8 & 58 & LEDA database, \citet{Eymeren2010}\\
$D$ / Mpc & 13.8 & 2.44 & \citet{Marlowe_1995,Karachentsev_2006} \\
$M_\text{B}$ / mag & $-17.68$ & $-16.27$& LEDA database \\
log($M_\text{HI}$ / M$_\odot$) & -- & 8.26 & \citet{Zastrow_2013,Koribalski_2018} \\
12 + log(O/H) & 8.3 & 8.1 & \citet{Hadfield_2006,Crowther_2009}\\
log[SFR / (M$_\odot$ yr$^{-1}$)] & $-0.15$ & $-1.12$ & \citet{Marasco_2023,Crowther_2009}\\
log[H$\alpha$ / (erg s$^{-1}$ cm$^{-2}$)] & $-11.46 $ & $-11.14$ &  \citet{GildePaz_2003,Crowther_2009}\\
$F_\text{HI}$ / (Jy km s$^{-1}$) & -- & 103.5 & \citet{Koribalski_2018}  \\
$S_{12\upmu \text{m}}$ / Jy & 0.31 & 0.27 & \citet{Condon_2021}\\
$S_{25\upmu \text{m}}$ / Jy & 0.74 & 1.27 & \citet{Condon_2021}\\
$S_{60\upmu \text{m}}$ / Jy & 5.33 & 8.82 & \citet{Condon_2021}\\
$S_{100\upmu \text{m}}$ / Jy & 6.67 & 11.38 & \citet{Condon_2021}\\
$^\dagger S_{0.2-2\,\text{keV}}$ / ($10^{-13}$ erg s$^{-1}$ cm$^{-2}$) & 1.902 & 0.70& SWIFT-XRT, \citet{Burrows_2005}\\
$^\dagger S_{2-12\,\text{keV}}$ / ($10^{-13}$ erg s$^{-1}$ cm$^{-2}$) & 54.89 & 33.67& SWIFT-XRT, \citet{Burrows_2005}\\
$^\dagger S_{20-40\,\text{keV}}$ / ($10^{-13}$ erg s$^{-1}$ cm$^{-2}$) & 31.72 & 32.64 & Integral, \citet{Walter_2003} \\
$^\dagger S_{40-60\,\text{keV}}$ / ($10^{-13}$ erg s$^{-1}$ cm$^{-2}$) & 27.48 & 28.63 & Integral, \citet{Walter_2003} \\
$^\dagger S_{60-100\,\text{keV}}$ / $(10^{-13}$ erg s$^{-1}$ cm$^{-2}$) & 28.42 & 29.88 & Integral, \citet{Walter_2003} \\
\bottomrule
\end{tabular}
\label{basics}
\begin{tablenotes}
\footnotesize
\item[$\dagger$] The X-ray flux are taken from \url{http://xmmuls.esac.esa.int/upperlimitserver/} and only represent an upper limit.
\end{tablenotes}
\end{threeparttable}
\end{table*}

\subsection{Selection of targets}
NGC\,3125 and IC\,4662 are two of the brightest dwarf galaxies in the IRAS RBGS \citep[][]{Sanders_2003}, which is the most complete sample of optically bright star-forming galaxies in the local Universe (see Table~\ref{Tab: properties} for a list of the properties of these two dwarfs).

NGC\,3125, an irregular dwarf galaxy, is uniquely showing the presence of kiloparsec-scale super-bubbles and filaments which tend to be oriented along the galaxy minor axis out to several kiloparsec as traced by the H$\alpha$ emission.
Such a galaxy is expected to host a strong superwind capable of escaping the galaxy and ejecting a significant fraction of the newly synthesised heavy elements into the intra-galactic medium \citep{Kronberg1999}. Indeed, ground-based optical imaging revealed a complex of H$\alpha$ filaments and shells \citep{Marlowe_1995} that is impressive even when compared to famous starburst dwarfs such as NGC\,1569, which has a similar mass and star formation rate (SFR) surface density. 
An optical polarisation map of NGC\,3125 shows that the complex of H$\alpha$ filaments is in part a large-scale bipolar reflection nebula originating from the brightest inner feature of the central starburst region \citep{Alton1994}. 
It is assumed that the dust in the nebula, which is concentrated along the major axis of the galaxy, originated in the central starburst regions and was driven out into the galaxy by starburst winds \citep{Alton1994}. 
\citet{Stevens_2002} has already conducted a resolved observation of NGC\,3125 at 4.80 and 8.64\,GHz with the Australian Telescope Compact Array (ATCA), displaying distinctive radio emission, with multiple discrete emission regions coinciding with massive star clusters. These emission regions consist of synchrotron radiation and optically thin H{\sc ii} regions. The synchrotron emission typically exhibits a spectral index of approximately $-0.5$, while optically thin H{\sc ii} regions have a spectral index of $\alpha_\text{tot} = -0.1$ \citep{Stevens_2002}. 
NGC\,3125 is highly X-ray luminous \citep[$L_{\rm{X}} \approx 2 \times 10^{39}$\,erg\,s$^{-1}$ in the $0.1-2.5$\,keV waveband,][]{Stevens_2002}, primarily due to high-mass X-ray binaries associated with active star formation. There is no evidence to suggest the presence of an active galactic nucleus (AGN) \citep{Stevens_2002}. Although supernovae or supernova remnants (SNRs) can be highly luminous in X-rays and radio waves \citep[e.g.,][]{vanDyk_1993,Immler_2003} with a relatively short period. Consequently, it is probable that high-mass X-ray binaries are the primary sources of the observed X-ray emission.
\citet{Zastrow_2013} observed an optically thin ionisation cone in NGC\,3125, with a clear excess in [S{\sc iii}]/[S{\sc ii}], indicating that the gas is highly ionised, implying a harder ionising radiation field with more high-energy photons, likely from an AGN or young, hot stars and a deficiency of [S{\sc ii}]/H$\alpha$, supporting the presence of strong ionising radiation, in the north-east side of the cone. This behaviour is consistent with expectations for optically thin gas \citep{Pellegrini_2012}.

Our second target, IC\,4662, a dwarf galaxy with a large Wolf-Rayet star population indicating a high star formation rate surface density of $\log[\Sigma_\text{SFR}/(\text{ M}_\odot \text{ yr}^{-1} \text{ kpc}^{-2}$)] = -1.39  
\citep{Hunter_2001}, potentially serving as progenitors of Type\,Ic supernovae gamma ray bursts\citep[e.g.,][]{Hammer2006,Modjaz2008}.
\citet{Johnson_2002} discuss the optical and radio morphologies of IC\,4662 and note that the radio morphology is dominated by two main regions of emission with thermal (inverted) spectra with $\alpha_\text{tot} = +0.3 \pm 0.2$ and $+0.4 \pm 0.2$ between 4.86 and 8.6\,GHz, due to possible free--free absorption. 
The positive spectral index is indicative of optically thick free--free thermal emission, which could result from ionised free--free electrons produced by O-type stars and/or supernovae \citep{Johnson_2002}. The two massive star-formation regions in IC\,4662 consist of several star clusters with ages of around $4\times10^6$\,yr and masses of approximately $3\times10^5\,$M$_\odot$. The clusters have high excitation, sub-solar abundances, and high extinctions of around $20\textrm{--}25$\,mag with dust well-mixed with the emitting gas \citep{Gilbert2009}. At high frequencies, the thermal emission from H{\sc ii} regions is expected to become optically thin, however, \citet{Johnson_2002} suggest this should be consistent with the data presented in their paper as they note that the observed low-frequency spectral index values indicate that the optically thick part of the emission at lower frequencies is contributing significantly to the total observed flux density. 
\citet{Crowther_2009} focused on spatially locating Wolf-Rayet stars through optical and spectroscopic observations, revealing He\,{\sc ii} excess and H$\alpha$ emission regions associated with 13 WC and 28 WN stars.
The abundances of various chemical elements were determined in IC\,4662, including hydrogen, helium, oxygen, nitrogen, neon, argon, sulfur, and iron. The findings reveal a nearly constant distribution of these elements throughout the face of the two massive star-forming regions in IC\,4662 \citep{Hidalgo_2001}. 
The neutral and ionised gas kinematics are intricate to compare due to the distorted velocity field of the H{\sc i} gas. Notably, the H{\sc ii} region in the southern part exhibits abnormal morphology. Analysis of H{\sc i} data suggests the gas in this dwarf galaxy is likely expanding \citep{Eymeren2010}.

Thus, NGC\,3125 and IC\,4662 are excellent candidates to study the suspected magnetised outflows from dwarf starburst galaxies, which will contribute to our understanding of the influence of CRs on galactic properties as well as on their ability to drive galactic winds. This understanding will clarify the evolutionary paths of these galaxies and offer comprehensive insights into the feedback mechanisms that regulate the formation and evolution of starburst dwarf galaxies.

\subsection{Observations}
\label{observations}

\begin{table*}
\caption{Summary of the ATCA Radio Observations of NGC\,3125}
\begin{tabular}{lcccccc}
\toprule
Date             & 4. May 2023 & 6. May 2023 & 7. May 2023 & 4. June 2023   & 21. July 2023 & 8. August 2023 \\
\hline
Observation Time / min  & 720 & 210 &  720 &  720  & 90 & 90 \\
Configuration    & 1.5A       & 1.5A       & 1.5A       & 6D            & 6D           & 6D            \\
Central Frequency / MHz    &    2100        &     2100       &   2100         & 2100          &      2100        &    2100           \\
Bandwidth / MHz  &    2048        &     2048       &       2048     &  2048             &  2048            &  2048             \\
Flux Calibrator  & 1934-638   & 1934-638   & 0823-500   & 0823-500      & 1934-638     & 1934-638      \\
Phase Calibrator & 1015-314   & 1015-314   & 1015-314   & 1015-314      & 1015-314     & 1015-314   \\
\bottomrule
\end{tabular}
\label{obs_ngc}
\end{table*}

\begin{table*}
\caption{Summary of the ATCA Radio Observations of IC\,4662}
\begin{tabular}{lcccccc}
\toprule
Date             & 4. May 2023 & 5. May 2023 & 6. May 2023  & 7. May 2023 & 4. June 2023 & 28. January 2024 \\
\hline
Observation Time / min  & 330 & 420 & 420 &  330  & 330 & 420 \\
Configuration    & 1.5A       & 1.5A       & 1.5A       & 1.5A            & 6D   & EW367  \\
Central Frequency / MHz    &    2100        &     2100       &   2100         & 2100          &      2100      & 2100       \\
Bandwidth / MHz  &    2048        &     2048       &       2048     &  2048             &  2048  & 2048  \\
Flux Calibrator  & 1934-638   & 1934-638   & 1934-638   & 1934-638      & 1934-638   &  1934-638  \\
Phase Calibrator & 1814-637   & 1814-637  & 1814-637   & 1814-637    & 1814-637  & 1814-637   \\
\bottomrule
\end{tabular}
\label{obs_ic}
\end{table*}

The two dwarf galaxies NGC\,3125 and IC\,4662 were observed with the ATCA (project ID: C3531; PI: S. Taziaux). The observations were performed at 2.1\,GHz (16\,cm) between 4-May-2023 and 28-January-2024, with the array in the 1.5\,km (1.5A), 6\,km (6D) and (EW367) configurations. Observations of the target galaxies and a phase calibrator were alternated through the observing run, giving a total observing time of about 43\,h on NGC\,3125 and 37.5\,h on IC\,4662. The phase calibrators used in these observations were 1016–311 for NGC\,3125 and 1814-637 for IC\,4662, along with the standard ATCA primary flux calibrator, 1934–638, except for two measurement sets where we use the secondary flux calibrator 0823-500, which has been calibrated through the phase. A summary of the observations used in this paper is given in Tables~\ref{obs_ngc} and \ref{obs_ic}.

\subsection{Data reduction}
\label{datareduction}

We employed data reduction procedures based on the Multichannel Image Reconstruction Imaging Analysis and Display \citep[\texttt{miriad;}][]{miriad}. 
The data reduction followed a standard procedure. To mitigate the impact of radio frequency interference (RFI) during flux and phase calibration, we utilised the interactive flagging tool \texttt{bflag}. Additionally, automated flagging routines were applied with the task \texttt{pgflag} to address interference for the source. 
Any additional corrupted data identified during calibration was manually flagged. 
Initially, after cross-calibration and excluding channels with H{\sc i} emission, we employed iterative imaging and self-calibration \texttt{selfcal} (phase-only, frequency-independent, with a solution interval of 10\,min down to 1\,min) until image quality reached convergence. 
Multifrequency CLEANing \texttt{mfclean} was conducted utilising interactive masks around visible sources to minimise artefacts and flux scattering and applying detection thresholds to retain genuine emission only.
The continuum images for each galaxy were generated using Briggs weighting with $robust = 0$, resulting in resolutions, noise levels, and image sizes, as provided in Table\,\ref{results}. In the dataset observed on the 4-June-2023 for IC\,4662, data from antenna 6 has been excluded due to an error in correlation block 1-bcc16 on antenna 6, which could not be fixed during the observation. 
In order to identify whether there is a missing zero spacing problem, as the long baselines were used, simulations were conducted which revealed no issues.

For polarisation imaging, we individually imaged the Stokes-$Q$ and -$U$ parameters. Each channel of 1\,MHz was imaged and cleaned (using the final mask from the total power imaging) to mitigate bandwidth depolarisation effects. We then corrected each image for primary beam attenuation. To include also the lowest frequencies, in order to utilise the largest bandwidth possible, we convolved all our images to a circular beam of $15\arcsec$. This was performed for both galaxies. 
After imaging, we checked all individual $Q$- and $U$-images again and removed any image still showing artefacts to achieve optimal quality for the following analysis steps. 
The final Stokes $Q$ and $U$ images, detailed alongside the frequency range and the number of images used in Rotation Measure Synthesis in Table~\ref{results}, served as input for the algorithm described by \citet{Burn_1966}.
Within this approach, the data is transformed from frequency space into Faraday space via a Fast Fourier Transformation (FFT). This technique is mathematically similar to aperture synthesis imaging and known as Rotation Measure Synthesis \citep{Brentjens_2005}. The instrumental parameters are determined by the instrument’s frequency configuration, while the instrumental function is described by the Rotation Measure Transfer Function (RMTF).
The resolution in Faraday space, the maximum observable scale and the maximum observable Faraday depth can be taken from the Table \ref{results}.
The $Q$- and $U$-cubes have been sampled between $-2048\,\mathrm{rad}\,\mathrm{m}^{-2}$ and $2048\,\mathrm{rad}\,\mathrm{m}^{-2}$ with a step size of $8\,\mathrm{rad}\,\mathrm{m}^{-2}$.
We then calculated a polarised intensity cube, which is the absolute value of the complex polarised intensity $p = Q+iU$, where $Q$ and $U$ are the fluxes in the Faraday $Q$- and $U$-cubes. 

Due to the fact that the intensity can only be positive, the polarised intensity can be described by the complex conjugated $P=pp^*=\sqrt{Q^2+U^2}$ following a non-Gaussian statistic \citep{Wardle_1974}, which is contingent upon the distance from the pointing centre due to primary beam correction.
When deriving a polarised intensity map from RM-Synthesis cubes, it is observed that the bias is elevated compared to the theoretical expectation. 
This is attributed to the calculation of polarised intensity from the maximum value along the Faraday axis. 
In case, noise dominates along this axis, there is a higher probability of encountering a higher value \citep{Heald_2009_II}.
The extent of this deviation hinges on the length of the sampled Faraday axis and the Faraday resolution of the observation.
Consequently, an approximate method was adopted to mitigate this bias \citep{Adebahr_2017}.
To address this, a higher-order polynomial with a parabolic shape was fitted to the distance of each pixel from the pointing centre against the average value along the polarised intensity axis in the Faraday polarised intensity cube.
This was performed solely in emission-free regions, termed the background. 
The background image was then extended along the third axis and subtracted from the polarised intensity cube.
The polarised intensity map and the RM map were derived by fitting a parabola along the Faraday axis of the polarised intensity cube to values exceeding $5\sigma$ \citep{Heald_2009_II}. 
The peak value of this parabola yields the polarised flux density, while the position of the peak in the Faraday spectrum determines the RM value.
RM is an observable for the frequency of rotation of the electric field vector along the line-of-sight due to Faraday rotation. 
Positive values represent a magnetic field that is pointing towards us while negative ones describe the opposite.
The orientation of the electric field vectors was derived using the polarisation angle $\chi$, calculated as $\chi=\frac{1}{2}\tan^{-1}\frac{U}{Q}$.
Assuming that the RM and polarised intensity originate from the same region, the electric field vectors $\Vec{E}$ were derotated back to the intrinsic magnetic field using $\chi_0=\chi - \text{RM}\,\lambda^2+\frac{\pi}{2}$.

\subsection{MeerKAT archive data}
\begin{figure}
    \centering
    \includegraphics[width=\linewidth]{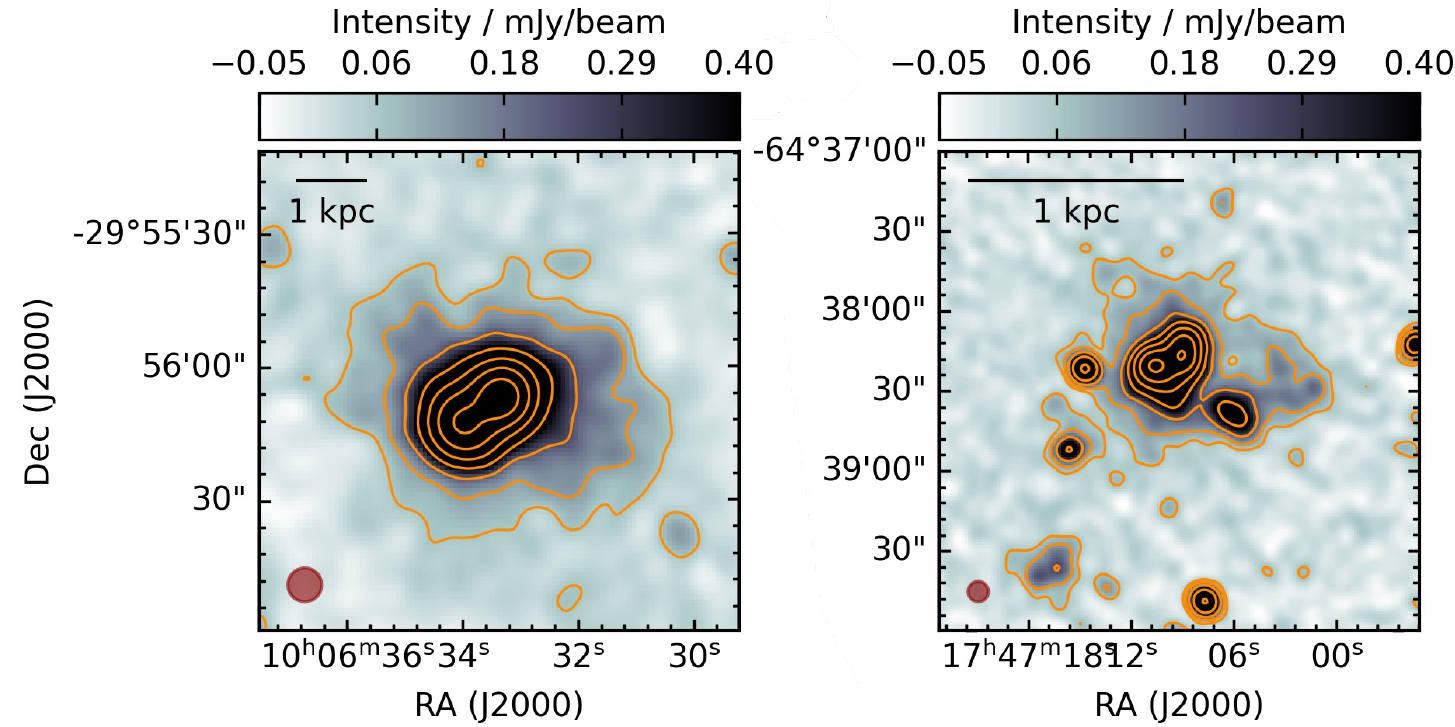}
    \caption{Total intensity emission at the central frequency of 1.28\,GHz with overlaid contours starting at $3\sigma$ and increase by factor of 2 ($\sigma = 20\,\upmu$Jy/beam). \textit{Left:} Total Intensity of NGC\,3125. \textit{Right:} Total Intensity of IC\,4662. 
    The circular beam of 7.6" appears in the lower left corner.}
    \label{meerkat_ti}
\end{figure}

The MeerKAT radio telescope consists of 64 dish antennas and observes in a frequency range of $856-1712$\,MHz. Its data is collected across 4096 spectral channels. The MeerKAT maps, taken from \citet{Condon_2021} and shown in Fig.~\ref{meerkat_ti}, are 15\,min snapshot observations at a central frequency of 1.28\,GHz. The data reduction, including amplitude, phase and polarisation calibration, imaging processing of the Stokes parameters and phase-only self calibration are done with \textit{Obit} \citep{Cotton_2008} at a \textit{robust briggs weighting} = $-1.5$. For more details about the data reduction see \citet{Condon_2021}.

\section{Properties of radio continuum polarimetry}
\label{radiocont}
\begin{figure*}
    \centering
    \includegraphics[width=0.85\linewidth]{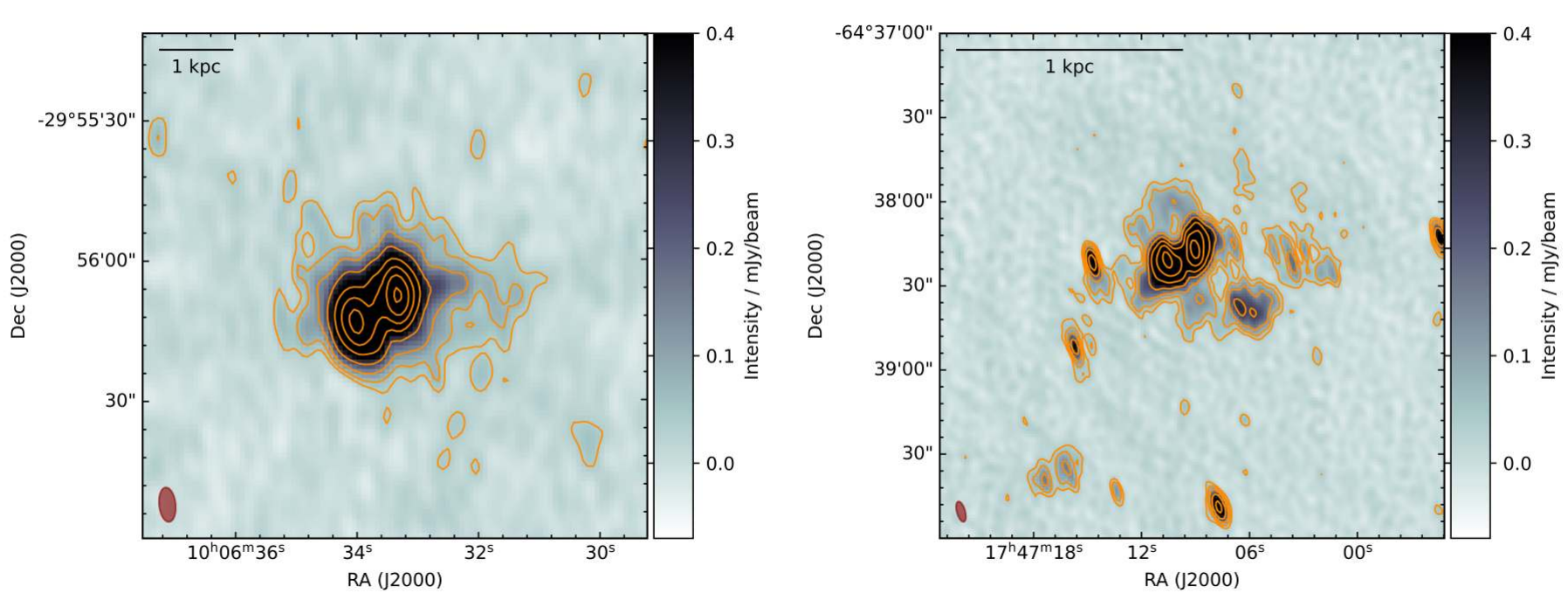}
    \caption{Total intensity emission at the central frequency of 2.1\,GHz with overlaid contours starting at $3\sigma$ and increase by factor of 2 ($\sigma = 11\,\upmu$Jy/beam). \textit{Left:} Total Intensity of NGC\,3125. \textit{Right:} Total Intensity of IC\,4662. 
    The beam of 7.3"$\times$3.5" for NGC\,3125 and 7.4"$\times$2.9" for IC\,4662 appears in the lower left corner and can also be taken from Table \ref{results}.}
    \label{TI}
\end{figure*}

\begin{figure*}
    \centering
    \includegraphics[width=0.85\linewidth]{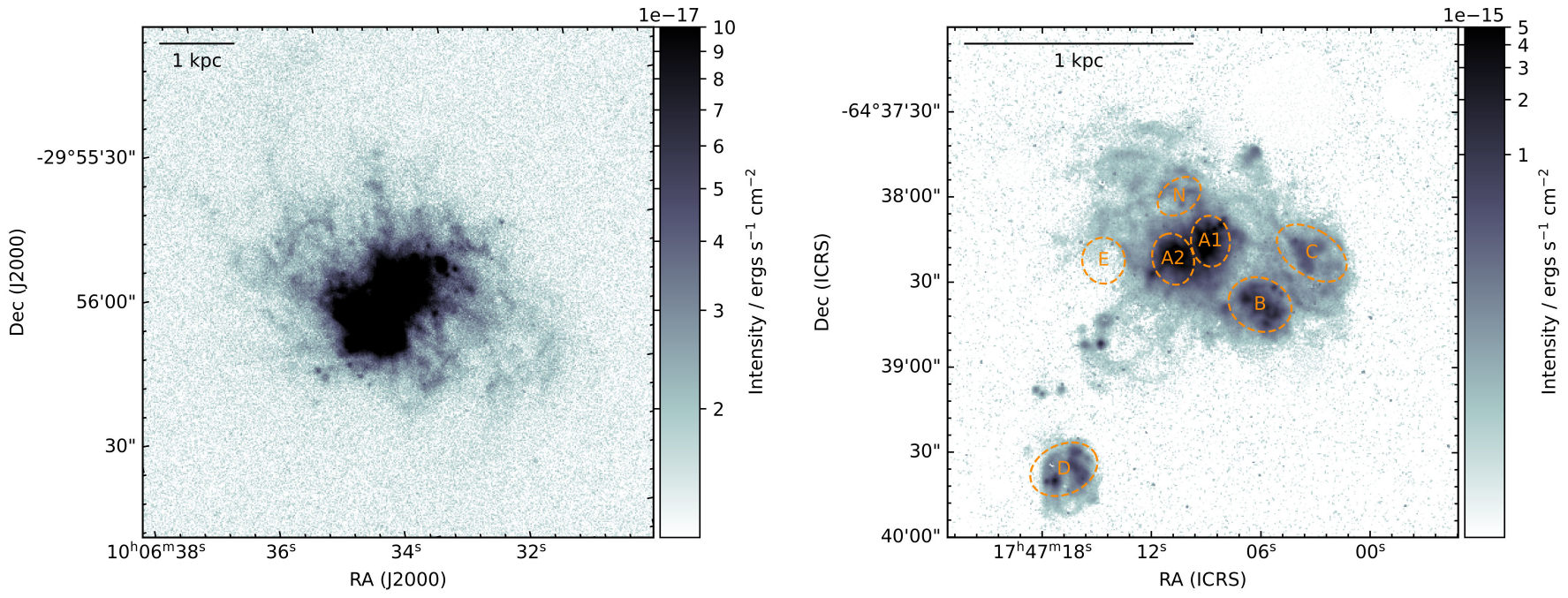}
    \caption{\textit{Left:} H$\alpha$ map of NGC\,3125 from \citet[][]{GildePaz_2003}. \textit{Right:} H$\alpha$ map of IC\,4662 from \citet{Hunter_2004}. The different regions are marked in orange.}
    \label{halpha}
\end{figure*}
The radio spectrum of galaxies is a combination of free-free emission and synchrotron radiation \citep[e.g.,][]{klein_2018}, which can be written as a combination of two power-laws in the optically thin regime.
The non-thermal synchrotron component has a spectral index ($S\propto \nu^{\alpha_\text{nth}}$) which is known to dominate at frequencies up to around $10\,$GHz, with a slope for SF regions of around $-0.7 \pm 0.03$ with little deviation \citep{Resmi_2021}.
By separating the synchrotron emission from the free–free emission, the total magnetic field strength can be estimated by assuming equipartition (see Appendix~\ref{MF}). Since free-free emission lacks polarisation and synchrotron emission is polarised, any polarised emission indicates synchrotron emission, providing insights into the magnetic field component (see Sect.~\ref{pi_sect} and Appendix~\ref{frm}).

\begin{figure*}
    \centering
    \includegraphics[width=0.85\textwidth]{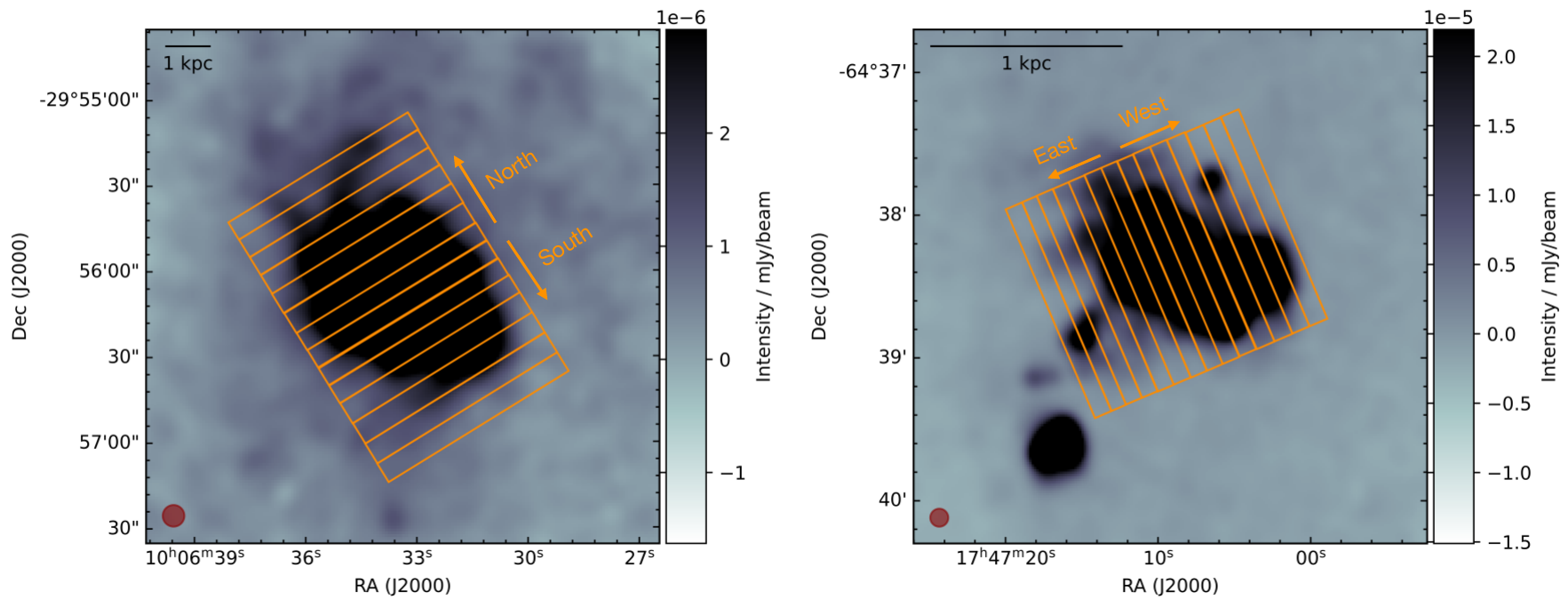}
    \caption{\textit{Left:} Thermal emission map of NGC\,3125 with overlaid boxes reaching from the south to the north. \textit{Right:} Thermal emission map of IC\,4662 with overlaid boxes reaching from the west to the east. The circular beam of $7.6\arcsec$ is shown in the left corner.}
    \label{box_maps}
\end{figure*}

\subsection{Total intensity}
\label{TI_sect}

Radio continuum emission has been observed from NGC\,3125 at a frequency of 2.1\,GHz, displayed in the left panel of Fig.~\ref{TI}.
The prominent radio emission is aligned with the optical main body of the galaxy, with its peak coinciding with the location of the giant H{\sc ii} region.
We detect both the bright central regions and the extended diffuse component.
There are two bright peaks of emission in the central regions of the galaxy, which are the two big SF knots.
The brightest peak is on the northeastern side of the disk, and the fainter, slightly elongated peak is on the southwestern.
We observe extended emission of NGC\,3125 perpendicular to the disk, resulting in a symmetric envelope of radio emission.

Using the same nomenclature as \citet{Johnson_2002} and \citet{Crowther_2009} for IC\,4662, the two central regions, A1 and A2, exhibit intense radio emission attributed to active star formation. The right panel of Fig.~\ref{TI} shows the total intensity of IC\,4662 and displays elongated structures, or filaments, extending from the central region to regions B, C, N and E. Additionally, several smaller, isolated regions of emission are scattered around the central area, with the most significant radio emission observed in the south-east (region D). 
This raises the question whether region D is an isolated H{\sc ii} region, a companion galaxy, or an extension of the main galaxy, which will be further discussed in Sect.~\ref{interaction}.

\subsection{Estimating the free--free emission}
\label{thermal_corr}

A separation of the non-thermal and thermal components of the radio emission is essential for the analysis of the spectral index (Sect.~\ref{spix}) and estimating the magnetic field strength (Appendix~\ref{MF}).
To obtain the thermal component, we utilise the free--free emission density function in relation to the H${\alpha}$ flux \citep[][equation\,3]{Deeg_1997}. 
For this purpose, we specifically use the H${\alpha}$ maps from \citet{GildePaz_2003} for NGC\,3125 and \citet{Hunter_2004} for IC\,4662 (see Fig.~\ref{halpha}). It is important to note that these maps have not been adjusted to account for potential contamination from the [N{\sc ii}] doublet emission lines or corrected for dust extinction, which could introduce systematic uncertainties into the thermal flux estimates. However, \citet{GildePaz_2003} and \citet{Hunter_2004} corrected the H$\alpha$ flux of both galaxies, NGC\,3125 and IC\,4662, for both internal and galactic extinction.

In the later stages of the analysis, we encountered some uncertainties regarding the reliability of the thermal/non-thermal separation. Consequently, the thermal and non-thermal fractions have been included in the Appendices~\ref{NT} and ~\ref{tf_sect}.
In the left panels of Fig.~\ref{syn} in Appendix~\ref{NT}, the non-thermal emission of NGC\,3125 extends beyond the core, forming a synchrotron halo around the galaxy. 
The radio envelope extends along the minor axis of the galaxy. The thermal contribution reveals a peak of 17.5\% to 22.5\% at 2.1\,GHz in the galaxy centre, with a rapid decline further out, shown in the left panels of Fig.~\ref{tf} in Appendix~\ref{tf_sect}. The mean thermal fraction of NGC\,3125 is estimated to be 11.08\,\% at 1.28\,GHz and 14.87\,\% at 2.1\,GHz.

In the right panels of Fig.~\ref{syn} in Appendix~\ref{NT}, we show the non-thermal emission from IC\,4662, which exhibits pronounced peaks corresponding to the bright star-forming regions A1 and A2. Additionally, there is a diffuse emission extending to the north toward region N and to the south, linking to the galaxy’s tail (regions B and C). The overall emission from this dwarf galaxy is predominantly thermal, as evidenced by the thermal fraction shown in the right panels of Fig.~\ref{tf} in Appendix~\ref{tf_sect}. The mean thermal fraction of the main body of IC\,4662 is estimated to be 36.06\,\% at 1.28\,GHz and 46.47\,\% at 2.1\,GHz. The thermal fraction reaches its maximum value of approximately 70\,\% at 2.1\,GHz in both the main body and the southern region, which corresponds to H{\sc ii} region D. In this region D, the mean thermal fraction is 48.52\,\% at 1.28\,GHz and 45.57\,\% at 2.1\,GHz.

To estimate the uncertainties in the non-thermal emission, we start with a $10\,\%$ uncertainty in the total flux density, coming from the telescopes. Additional errors arise from the thermal correction, due to electron temperature variations $T_\text{e} = 10000 \pm 3000$\,K  and a $10\,\%$ uncertainty in the treatment of the [N{\sc ii}] line flux relative to H$\alpha$ measurements \citep{Vargas_2018}. Combining these factors, we estimate a total uncertainty of approximately $20\,\%$ in the final non-thermal flux density measurement. Note that this does not include systematic uncertainties and may be a gross underestimate.

\subsection{Box-integrated intensity profiles}
\label{box}

\begin{figure*}
\centering
    \includegraphics[width=0.85\textwidth]{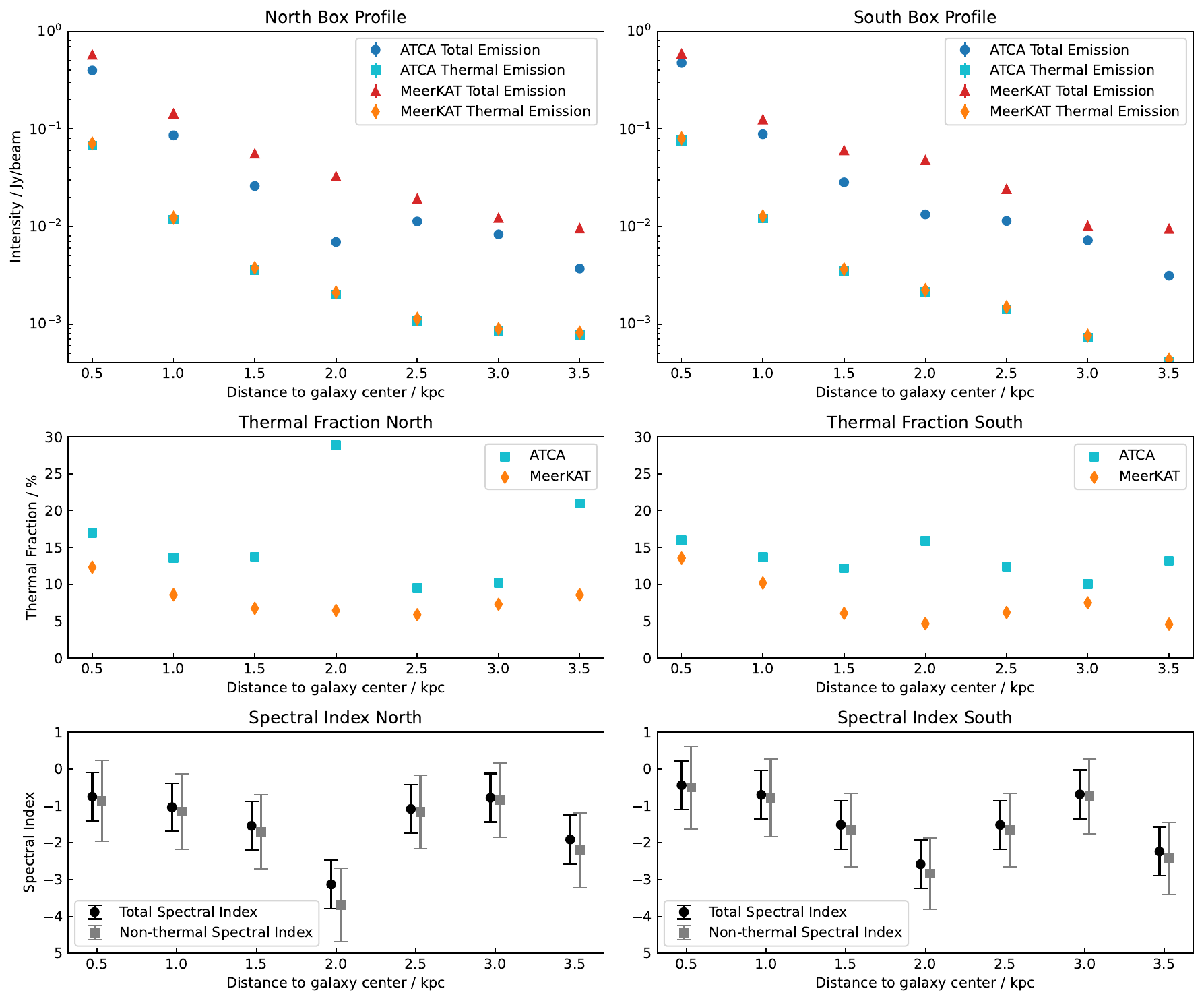}
    \caption{\textit{Upper Panel:} Box-integrated radio continuum profiles of NGC\,3125, separated into northern (left) and southern (right) sections, showing the total and thermal emission at 1.28\,GHz (MeerKAT) and 2.1\,GHz (ATCA). \textit{Middle Panel:} Thermal fraction profiles of NGC\,3125 for the northern (left) and southern (right) sections. \textit{Bottom Panel:} Comparison of non-thermal and total spectral indices for the northern (left) and southern (right) sections of NGC\,3125. Spectral index data points on the x-axis are slightly offset to improve differentiation between close values.
    }
    \label{ngc_box}
\end{figure*}

\begin{figure*}
\centering
    \includegraphics[width=0.85\textwidth]{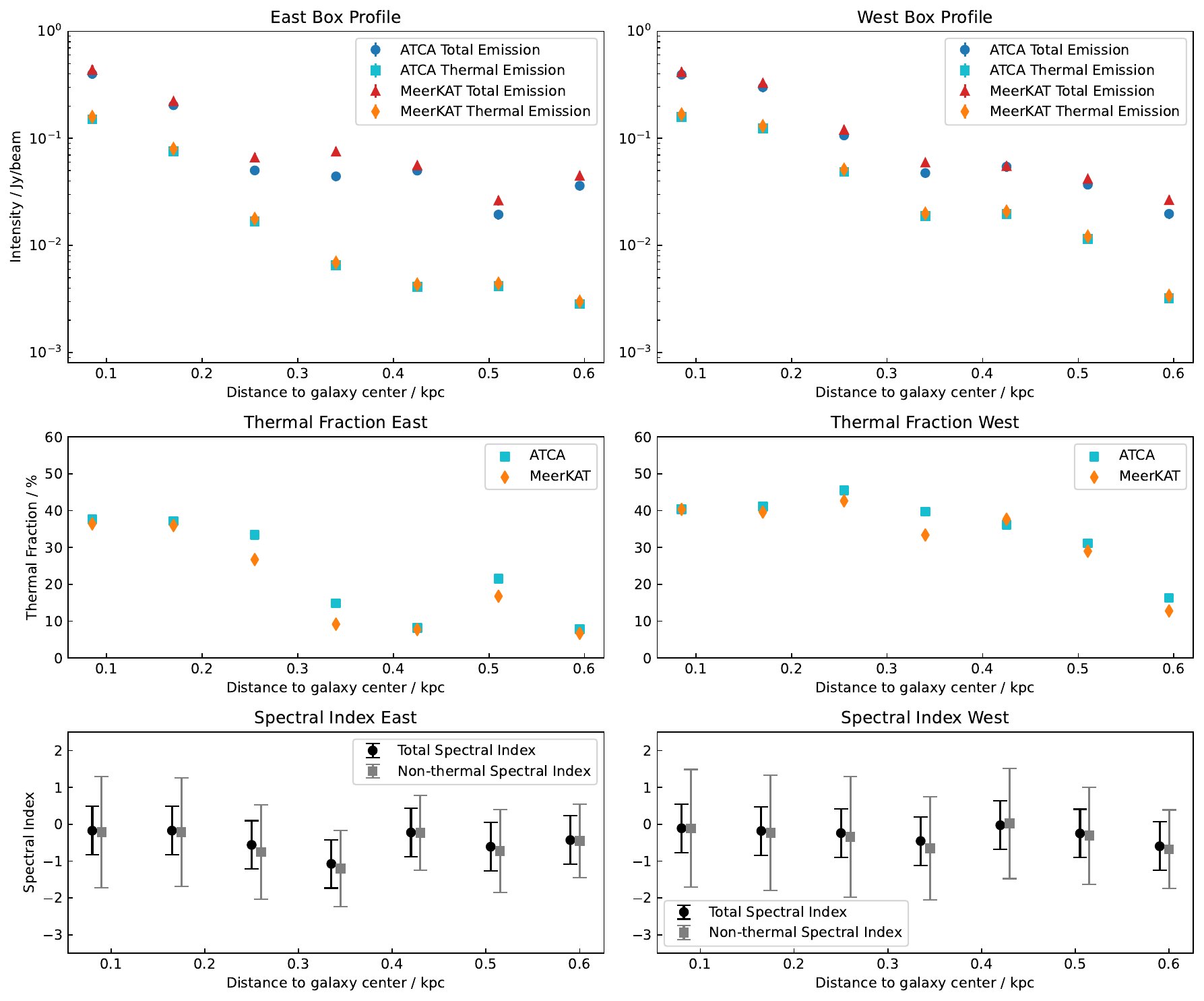}
    \caption{\textit{Upper Panel:} Box-integrated radio continuum profiles of IC\,4662, separated into eastern (left) and western (right) sections, showing the total and thermal emission at 1.28\,GHz (MeerKAT) and 2.1\,GHz (ATCA). \textit{Middle Panel:} Thermal fraction profiles of IC\,4662 for the eastern (left) and western (right) sections. \textit{Bottom Panel:} Comparison of non-thermal and total spectral indices for the eastern (left) and western (right) sections of IC\,4662. Spectral index data points on the x-axis are slightly offset to improve differentiation between close values.}
    \label{ic_box}
\end{figure*}

Spectral index derived from observed total intensity radio continuum emissions serve as crucial analysis tools for explaining the underlying emission mechanisms, and energy loss/gain processes of the synchrotron emitting CREs.
Furthermore, analysis of observed spectral indices provide insights into whether thermal or synchrotron emissions dominate within a specific region.

However, it is important to recognise that spectral index maps are susceptible to significant distortion due to minor systematic fluctuations in flux density. For instance, in the case of a frequency range between $1.28$\,GHz and $2.1$\,GHz, even a marginal $5\,\%$ fluctuation in flux density can cause the spectral index to change by as much as $\pm 0.104$, depending on whether the flux density increases or decreases. 

To assess the validity of the thermal/non-thermal separation and extend the study of radio continuum emission into the galactic halo, we utilise a box integration method to average data over large regions of NGC\,3125 and IC\,4662. We use one strip containing 14 boxes with the FWHM of our synthesised beam ($7.6"$) as box height for each galaxy, determined by the spatial extent necessary to include the halo region fully. This approach allows us to differentiate between the central star-forming regions, the outskirts, and the more distant outer regions. 
The total absolute flux density error arises from multiple sources, including background noise level and calibration uncertainties. Background noise increases with the number of beams within the integration area, contributing to the overall uncertainty \citep{Klein_1981}. However, in our analysis, the dominant source of error is the calibration uncertainty, which is estimated to be $10\,\%$.
For NGC\,3125, we separate the northern and southern sections, whereas for IC\,4662, we distinguish the western and eastern sections, using the optical major axis as the dividing line (see Fig.~\ref{box_maps}).

Fig.~\ref{ngc_box} (upper panel) shows the integrated total and thermal radio emissions for various regions of NGC\,3125, with separate analyses for the northern and southern sections. NGC\,3125 shows extended radio continuum emission up to a distance of 2\,kpc from the galaxy’s centre. In the northern section, it is noteworthy that there is a lack of detectable radio emission at 2.1\,GHz, while total emission is still discernible at a lower frequency of 1.28\,GHz.
We further extend the profile up to 3.5\,kpc, where thermal emission remains detectable (see left panel of Fig.~\ref{box_maps}). This extended range indicates that thermal processes are still active, even as radio continuum emission diminishes. Within this northern region, the morphology of the feature might indicate an outflow, prominently visible in the thermal emission maps. This outflow is less apparent in the total intensity map, where it falls below the $3\sigma$ sensitivity threshold, suggesting that the feature might become more distinguishable with increased observational sensitivity. 
At 2.1\,GHz, we observe an increase in the thermal fraction at 2\,kpc, which can be attributed to the absence of detectable radio continuum emission in this region, despite ongoing thermal emission.
Furthermore, the total and non-thermal spectral indices are found to be closely aligned, indicating similar contributions from different emission mechanisms.
Beyond the 2\,kpc threshold, an increase in the spectral index is detected, which may serve as evidence for the weak total radio continuum emission associated with the outflow arm in the northern section of the galaxy. This suggests a potential link between the outflow dynamics and the radio emissions. In addition, the southern part of NGC\,3125 exhibits a flatter spectrum compared to the northern section, which can also be observed in the left panel of Fig.~\ref{si_sc_ngc}, highlighting distinct differences in the physical conditions and processes occurring in these two regions of the galaxy. This disparity may reflect variations in star formation activity, magnetic field configurations, or different contributions from various emission processes across the galaxy.

\begin{table*}
\caption{Summary of the integrated total radio continuum flux density  values and its error for NGC\,3125 and IC\,4662}
\centering
\begin{threeparttable}
\begin{tabular}{lccc}
\toprule
$\nu$ / MHz  & S$_\text{tot, NGC\,3125}$ / mJy & S$_\text{tot, IC\,4662}$ / mJy & Reference   \\
\midrule
87.7 & $18 \pm 5.3$ & -- & \cite{gleam} \\
118.4 & $21 \pm 6.8$ & -- & \cite{gleam} \\
154.2 & $25 \pm 5.5$ & -- & \cite{gleam} \\
887 & $22.98 \pm 2.31$ & $29.57 \pm 2.96$ &  \cite{racs} \\
1280 & $25.0 \pm 2.50$ & $34.06 \pm 3.4$ & \cite{Condon_2021} \\
1460 & $22.88 \pm 2.28$ & $31.86 \pm 3.18$ & this work \\
1716 & $21.35 \pm 2.13$ & $31.87 \pm 3.18$ & this work \\
1972 & $20.13 \pm 2.01$ & $30.69 \pm 3.07$ & this work \\
2228 & $17.96 \pm 1.79$ & $28.99 \pm 2.89$ & this work \\
2484 & $17.08 \pm 1.70$ & $26.93 \pm 2.69$ & this work \\
2740 & $^\dagger14.64 \pm 1.46$ & $26.92 \pm 2.69$ & this work \\
4800 & $10.2 \pm 1.02$ & $8.2 \pm 0.93$ & \cite{Stevens_2002}, \cite{Johnson_2002} \\
8640 & $7.5 \pm 0.75$ & $9.7 \pm 1.11$ & \cite{Stevens_2002}, \cite{Johnson_2002} \\
\bottomrule
\end{tabular}
\begin{tablenotes}
\footnotesize
\item[$\dagger$] There may be some flux loss due to the use of only long baselines.
\end{tablenotes}
\end{threeparttable}
\label{int_sed}
\end{table*}

\begin{figure*}
    \centering
    \includegraphics[width=\linewidth]{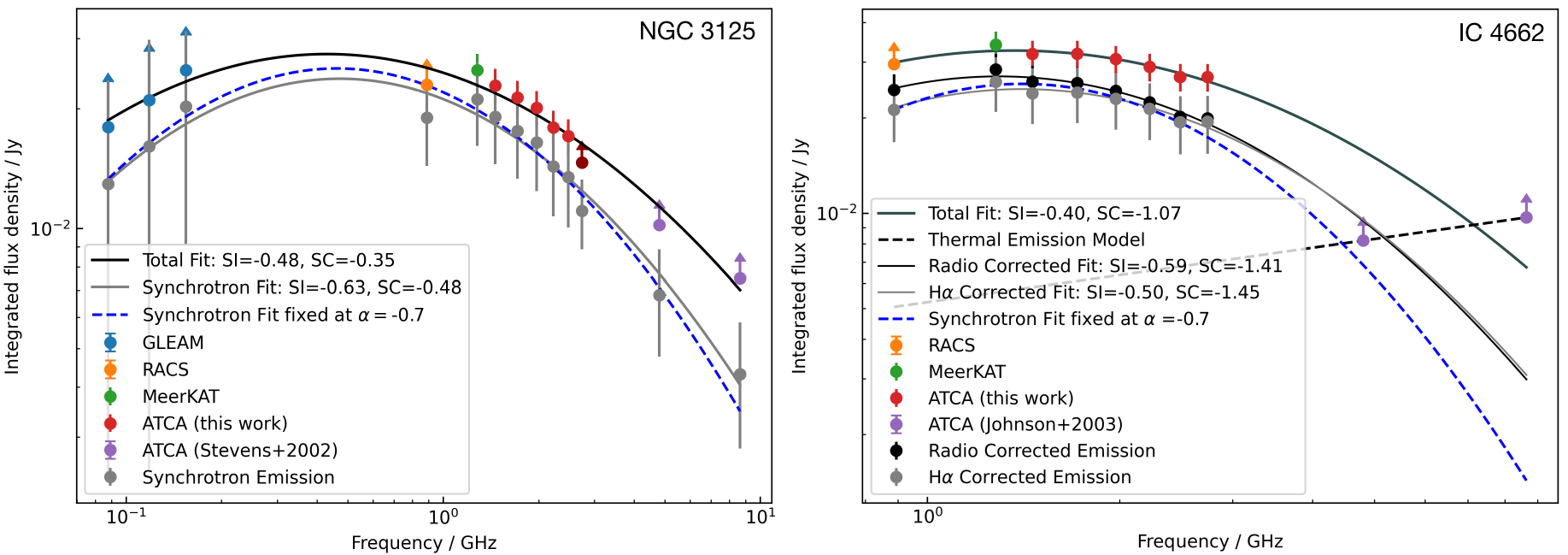}
    \caption{Spectral energy distribution of NGC\,3125 (\textit{left panel}) and IC\,4662 (\textit{right panel}). The plot displays total intensity data with different observations: GLEAM in blue, RACS in orange, MeerKAT in green, ATCA data from this study in red, and high-frequency data from ATCA for NGC\,3125 \citep{Stevens_2002} and for IC\,4662 \citep{Johnson_2002} in violet. As GLEAM and RACS are survey data, we only use lower limits as we expect more flux to be there. Additionally, for NGC\,3125 there may be some flux loss at 2740\,MHz, therefore the data point, marked in dark red and showed as lower limit, does not contribute to the fit. Thermally corrected emissions are shown in gray. Additionally, the polynomial fits from \citet{Perley_2017} are provided, indicating the spectral index $\alpha_0$ (SI) and spectral curvature $\beta$ (SC) of the source. A standard synchrotron spectrum fit with a spectral index of $\alpha = -0.7$ is shown in blue. For IC\,4662, we additionally present a thermal radio corrected emission in black by assuming only thermal emission above 4.86\,GHz and using the black dashed line as new thermal emission model. The 4.8 and 8.64\,GHz data points for IC 4662 are excluded from the fit.
    }
    \label{sed_int}
\end{figure*}

In Fig.~\ref{ic_box}, we observe the significant part of thermal emission in the central regions of IC\,4662, with thermal fractions reaching values up to $40\,\%$. The total and non-thermal spectral indices in this galaxy show minimal distinction, resulting in a generally flat spectral profile across the entire galaxy. Notably, the western section, which encompasses regions A1, B, C, and part of region N, exhibits a markedly flatter spectrum compared to the eastern region, which includes areas A2, E, and the other half of region N.
The pronounced flatness of the spectrum in IC\,4662, as confirmed by multiple analyses through this study, may be attributed to the combined effects of free--free emission and an extreme degree of low-frequency absorption \citep{Werhahn_2021}. This observation suggests that the traditional methodologies for estimating free-free emission, commonly derived from H$\alpha$, may not be applicable or effective in this particular dwarf galaxy.

\subsection{Integrated spectral index}
\label{spix}

\begin{figure*}
    \centering
    \includegraphics[width=\linewidth]{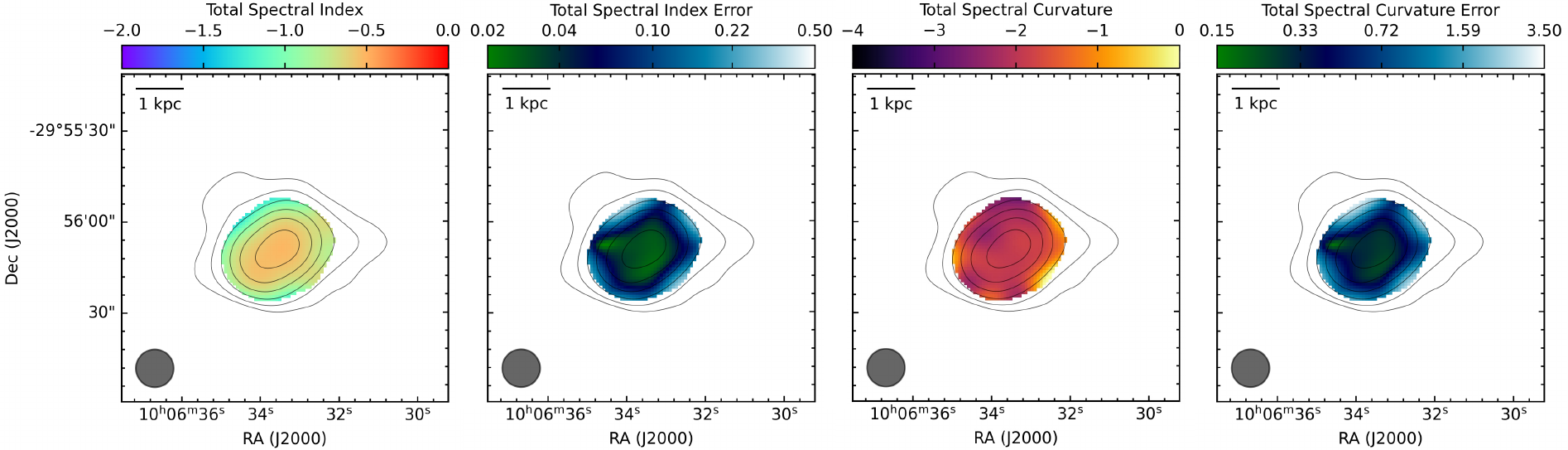}
    \caption{ 
    \textit{Left:} Fitted in-band total spectral index and its error of NGC\,3125 between the frequency ranges 1460\,MHz and 2740\,MHz. \textit{Right:}  Fitted in-band total spectral curvature and its error of NGC\,3125 between the frequency ranges 1460\,MHz and 2740\,MHz. The contours show the 2228\,MHz total radio emission at (5, 10, 20, 40, 80, 160, 320) $\times \,37.5\,\upmu$Jy/beam. The circular beam of $12.5"$ is shown in the left corner.}
    \label{si_sc_ngc}
\end{figure*}

\begin{figure*}
    \centering
    \includegraphics[width=\linewidth]{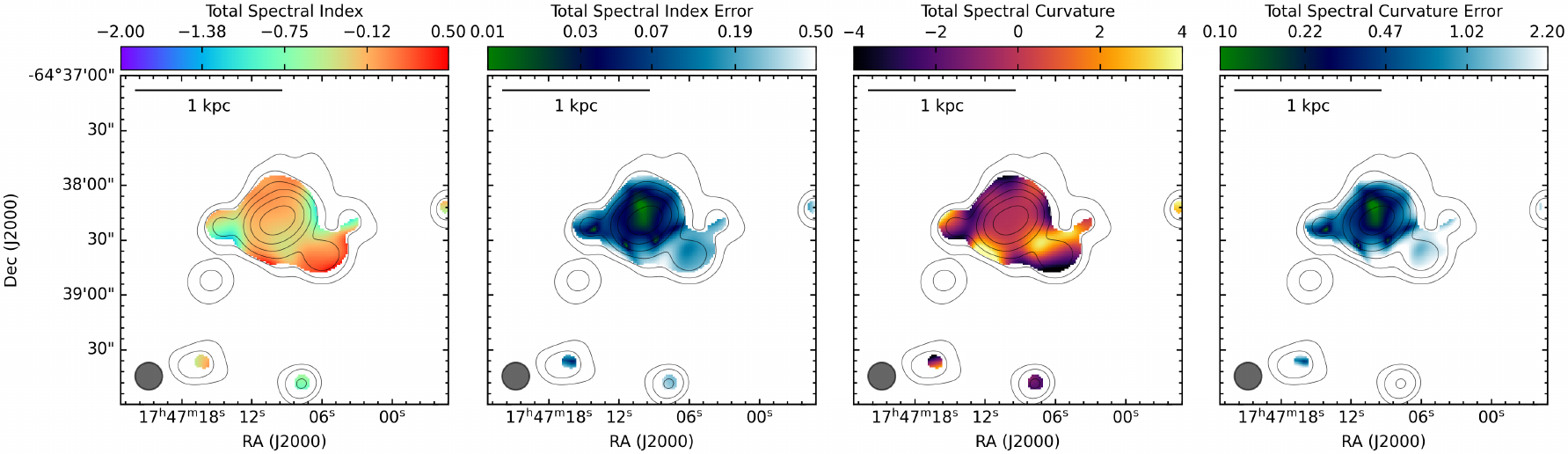}
    \caption{ 
    \textit{Left:} Fitted in-band total spectral index and its error of IC\,4662 between the frequency ranges 1460\,MHz and 2740\,MHz. \textit{Right:}  Fitted in-band total spectral curvature and its error of IC\,4662 between the frequency ranges 1460\,MHz and 2740\,MHz. The contours show the 2228\,MHz total radio emission at (5, 10, 20, 40, 80, 160, 320) $\times \,38\,\upmu$Jy/beam. The circular beam of $15"$ is shown in the left corner.}
    \label{si_sc_ic}
\end{figure*}

Since we observe no significant difference between the total and non-thermal spectral index in the box integrated profiles, and to minimise systematic errors, we present the spectral index analysis based on the total uncorrected flux densities. A corresponding analysis was also performed using non-thermal emission maps, with the results provided in Appendix~\ref{spix_nth}.

In order to estimate the accuracy of the spectral index, the data set was divided into eight segments, each with a bandwidth of 256\,MHz.
Due to high noise levels, we excluded the two edge bins at 1205\,MHz and 2996\,MHz. 
We then fitted a model spectrum based on a polynomial equation \citep{Perley_2017}

\begin{equation}
\log S_{\nu} = \log S_0 + \alpha_0 \log\left(\frac{\nu}{\nu_0}\right) + \beta \left[\log\left(\frac{\nu}{\nu_0}\right)\right]^2.
\label{perleybutlereq}
\end{equation}

where $S_\nu$ is the flux density at a given frequency $\nu$ (in GHz), $S_0$ is the flux density at the reference frequency $\nu_0= 2.1$\,GHz, $\alpha_0$ represents the polynomial spectral index at the reference frequency $\nu_0$, and $\beta$ is the spectral curvature.
We begin with this purely phenomenological model to investigate the general shape of the spectrum without bias towards any specific physical mechanism. For the integrated spectra, we incorporate additional data from several sources: GLEAM \citep{gleam}, RACS \citep{racs}, MeerKAT \citep{Condon_2021}, and high-frequency observations from ATCA for NGC\,3125 \citep{Stevens_2002} and for IC\,4662 \citep{Johnson_2002}. These data are also detailed in Table~\ref{int_sed}. 

In the left panel of Fig.~\ref{sed_int}, we present the integrated spectral energy distribution of NGC\,3125, as derived from the data in Table \ref{int_sed}. At frequencies above 4.86\,GHz, we have used the integrated values from all star-forming regions reported by \citet{Stevens_2002} as lower limits. The observed total spectral index is $-0.48\pm 0.03$, with a spectral curvature of $-0.35\pm 0.04$. The non-thermal spectral index, after accounting for H$\alpha$ emission, is $-0.63 \pm 0.05$, while the corresponding spectral curvature is $-0.48 \pm 0.05$. 
However, for NGC\,3125, there may be some flux loss at 2740\,MHz (the final data point of the splitted data) due to the use of only long baselines. This effect is noticeable, so the 2740\,MHz point should be treated as a lower limit.

We perform model comparisons using the Akaike Information Criterion (AIC) and the Residual Sum of Squares (RSS) to assess whether including curvature improves model performance. To do so, we fitted the model both with and without the curvature term. Summarised, AIC helps select the best model by penalizing complexity and avoiding overfitting and RSS measures how well the model fits the data, for both the lower, the better. Here, we consider the full spectrum in one case and only the ATCA data (excluding 2.7\,GHz) along with MeerKAT in the other.
For the full spectrum, the RSS is significantly reduced when including curvature ($1.71 \times 10^{-5}$ vs. $7.42 \times 10^{-4}$), and the AIC confirms a better fit with curvature ($-6.95$ vs. $25.03$). However, when considering only the ATCA and MeerKAT data, the RSS is slightly lower with curvature ($4.70 \times 10^{-7}$ vs. $5.51 \times 10^{-7}$), but the AIC increases ($-17.95$ vs. $-18.75$), indicating that the additional curvature parameter is not justified in this limited frequency range. Overall, while curvature improves the fit across the entire spectrum, a straight-line model is sufficient for the ATCA and MeerKAT data alone. 
Furthermore GLEAM and RACS, as survey data, are expected to show slightly higher integrated flux densities. A linear regression based on MeerKAT and ATCA data, extrapolated to 87\,MHz (GLEAM), predicts an integrated flux density of 32\,mJy, whereas the observed value is 18\,mJy, approximately 50\,\% lower. Additionally, within the GLEAM survey, the integrated flux density decreases with decreasing frequency, which cannot be attributed to missing coverage. 
This suggests that detecting spectral curvature requires including low-frequency data, also shown in \citet[]{Stein_2023}, \citet{Gajovic_2024} and \citet{Gajovic_2025}.

To compute an in-band spatial distributed spectral index, we split our ATCA dataset as described above, convolved it to the same resolution of a $12.5\arcsec$ circular beam, clipped each at $5$ times the rms noise level, and plotted the fitted total spectral index  and spectral curvature using the \citet{Perley_2017} polynomial equation, as shown in the left and right panel of Fig.~\ref{si_sc_ngc}. We find a total spectral index of $-0.64 \pm 0.03$ in the galactic disk and decreasing to $-1.22 \pm 0.19$ towards the edges of the galaxy with a spectral curvature ranging from $-2.44\pm 0.34$ to $-3.35 \pm 0.62$. 

Note that, the polynomial spectral index $\alpha_0$ and 2-point spectral index $\alpha$ are not the same. From Eq. (\ref{perleybutlereq}), we get that $\alpha_0=\alpha-2\beta[\log(\nu/\nu_0)]$. In order to compare $\alpha_0$ and $\alpha$ between 1.28 and 2.1\,GHz, seen in Fig.~\ref{ngc_box} and ~\ref{ic_box}, the reference frequency $\nu_0$ should be chosen somewhere between 1.28 and 2.1\,GHz, depending on the spectral curvature $\beta$. 
 
In right panel of Fig.~\ref{sed_int}, we show the integrated spectral energy distribution of IC\,4662 from Table~\ref{int_sed}. At frequencies above 4.86\,GHz, we have used the integrated values of all the regions provided by \citet{Johnson_2002} in their study, shown as lower limits but excluded from the fit. First, we observe a total spectral index of $-0.39 \pm 0.06$ with a spectral curvature of $-1.07\pm 0.19$. To estimate the non-thermal spectral index, the contribution from the thermal emission is subtracted using the integrated value of H$\alpha$ from Table~\ref{basics} \citep[][Eq.3]{Deeg_1997}. The non-thermal spectral index and spectral curvature are $-0.50 \pm 0.07$ and $-1.45 \pm 0.26$, respectively. 
Additionally, assuming that only free-free emission dominates at higher frequencies, we modeled the thermal emission using data at 4.8 and 8.64\,GHz. By subtracting this thermal contribution from the total intensity, we derived a non-thermal spectral index of $-0.59 \pm 0.07$ and a spectral curvature of $-1.41 \pm 0.23$.
For these latter two models, which include the H$\alpha$ corrected emission and the radio corrected emission, we did not incorporate the values provided by \citet{Johnson_2002} when calculating the spectral index and its spectral curvature.

Comparing the total and non-thermal spectral indices from the integrated spectrum with the two-point spectral index presented in Table~\ref{results}, a discrepancy is observed. This arises because the polynomial spectral index and the two-point spectral index are not directly comparable, as spectral curvature must be considered. Additionally, the two-point spectral index between the central frequencies of 1.28 and 2.1\,GHz represents only a narrow portion of the overall spectrum.

To compute the distribution in-band spectral index, we split our ATCA dataset as described, as shown in the middle and right panel of Fig.~\ref{si_sc_ic}. The central region remains flat or shows a positive spectrum, indicating a dominance of thermal emission and a reduced contribution from synchrotron radiation. In the core of the galaxy, the spectral curvature shows a variation between $-1.5$ and $1.5$.  
The positive spectral indices observed in north region indicate a dominance of optically thick thermal emission, further discussed in Sect.~\ref{sp_flatness}. In the top right panel of Fig.~\ref{si_nt} in Appendix~\ref{spix_nth}, we present the fitted non-thermal spectral index. In the same regions, the non-thermal spectral index is found to be positive, which is impossible to reconcile with CREs that have been accelerated via diffusive shock acceleration at supernova remnants \citep{Marcowith_2016}. Hence, this suggests that the H$\alpha$-based thermal correction is insufficient and not applicable for this type of dwarf galaxy. Similar flat spectra, with even positive total spectral indices for IC\,4662, have also been observed at higher frequencies by \citet{Johnson_2002}.

\subsection{Polarised emission}
\label{pi_sect}

\begin{figure*}
    \centering
    \includegraphics[width=\linewidth]{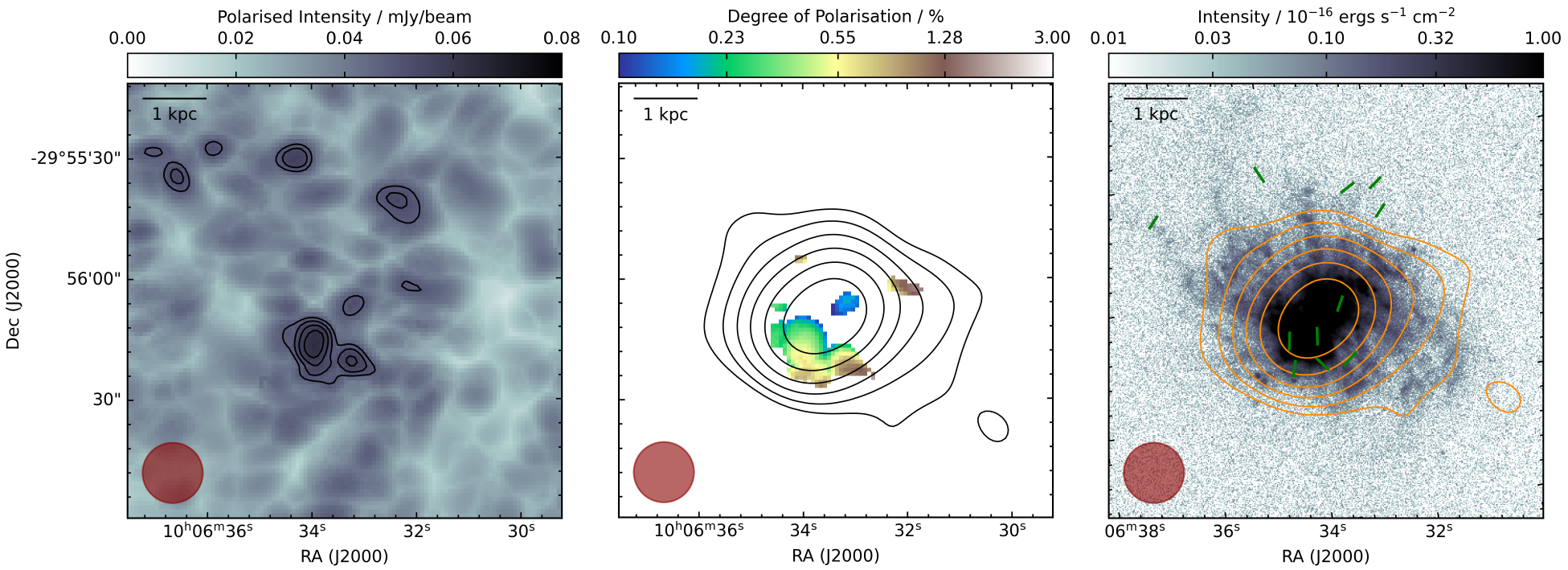}
    \caption{\textit{Left:} The polarised intensity of NGC\,3125. The contours show $5\sigma$ increasing with a factor of 1.5 plus the mean of the background ($\sigma = 1.54\,\upmu$Jy/beam and mean = $39.9\,\upmu$Jy/beam). \textit{Middle:} Degree of polarisation of NGC\,3125 with polarised emission clipped at $5\sigma$ ($\sigma = 1.54\,\upmu$Jy/beam) with subtracted background and total radio emission at 2.1\,GHz contours at $5\sigma$, increasing by a factor of 2 ($\sigma = 25\,\upmu$Jy/beam), overlaid. \textit{Right:} H$\alpha$ map of NGC\,3125 with non-thermal radio emission contours at $3\sigma$, increasing by a factor of 2 ($\sigma = 25\,\upmu$Jy/beam), overlaid. The  magnetic field orientations are shown in green and scale with the polarised intensity. For all three images, the circular beam of $15"$ is shown in the left corner.}
    \label{PI_ngc}
\end{figure*}

\begin{figure*}
    \centering
    \includegraphics[width=0.7\linewidth]{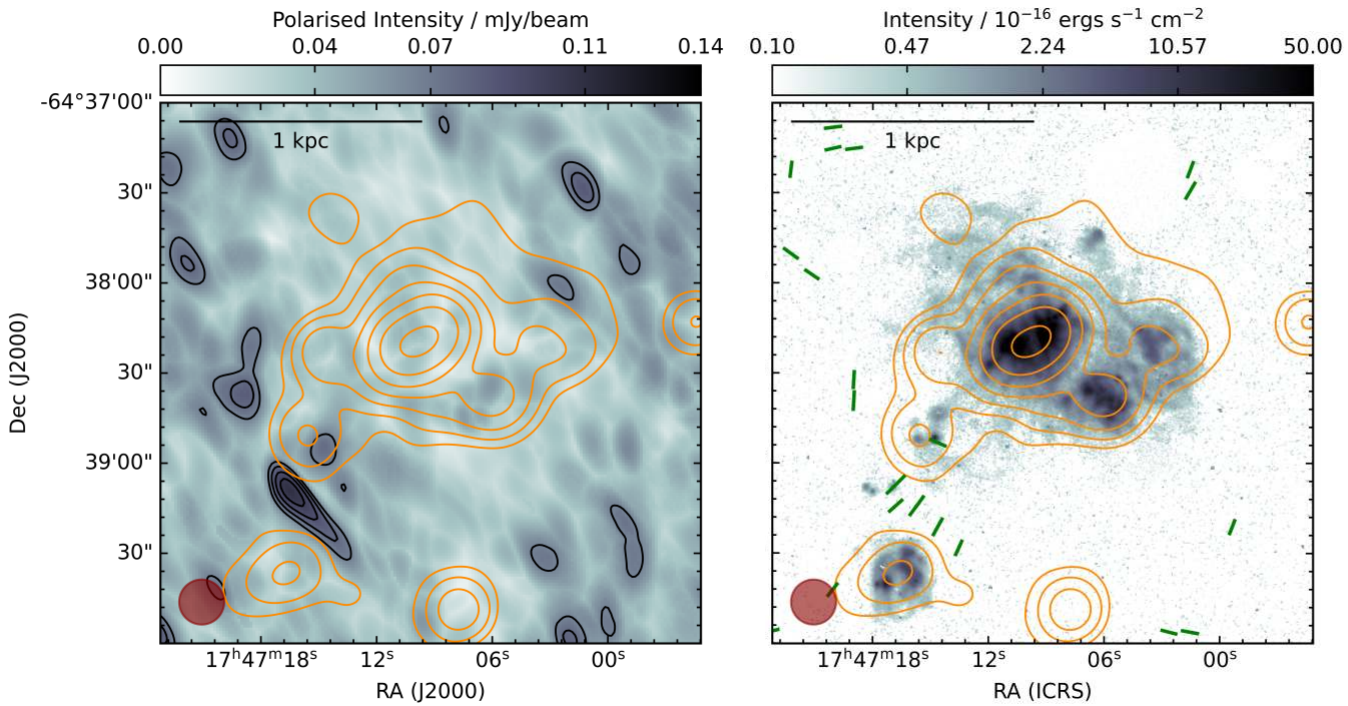}
    \caption{\textit{Left:} The polarised intensity of IC\,4662. The black contours show $5\sigma$  increasing with a factor of 1.5 plus the mean of the background  ($\sigma = 4.5\,\upmu$Jy/beam and mean = $42.2\,\upmu$Jy/beam) of the polarised emission. The orange contours show $5\sigma$  increasing with a factor of 2 ($\sigma = 25\,\upmu$Jy/beam) of the total radio intensity. \textit{Right:} H$\alpha$ map of IC\,4662 with total radio emission contours at $3\sigma$, increasing by a factor of 2 ($\sigma = 40\,\upmu$Jy/beam), overlaid. The  magnetic field orientations are shown in green. The circular beam of $15"$ is shown in the left corner.}
    \label{PI_ic}
\end{figure*}

The left panel of Fig.~\ref{PI_ngc} shows the distribution of polarised emission of NGC\,3125 across two distinct regions. 
Two regions were identified: the first is situated in the south of the two star-forming knots, while the second is located to the north-west of the central area.
In the southern region, the magnetic field orientations are oriented nearly perpendicular to the optical major axis of the galaxy, while moving towards the north-west, the magnetic field orientations progressively align more closely with the major axis of the galaxy. The integrated polarised emission of the galaxy is approximately $34.2\pm3.5\,\upmu$Jy.
In the right panel of Fig.~\ref{PI_ngc}, the H$\alpha$ emission is overlaid with magnetic field orientations. The filaments, as traced by the H$\alpha$ emission, do not show any detectable polarised emission. While filaments observed in H$\alpha$ can sometimes be associated with polarised emission \citep[e.g.,][]{kepley_role_2010, chyzy_magnetized_2016} due to aligned magnetic fields, in this case, no significant polarisation is observed within them.
We observe a degree of polarisation in these regions of $0.1\,\%$ up to $1.8\,\%$, seen in the middle panel of Fig.~\ref{PI_ngc}. The leakage of polarisation of the standard calibrator 1934-638 is $0.08\,\%$ \citep{Schnitzeler_2011}.

For IC\,4662, we do not detect any polarised emission (see Fig.~\ref{PI_ic}), likely due to turbulence obscuring the presence of ordered and/or anisotropic random magnetic fields, potentially resulting in depolarisation. The several processes of depolarisation, possibly happening this galaxy are discussed in the Sect.~\ref{depolarisation}. 
A clear detection of polarised emission at a mean weighted $\lambda^2$ of $0.0202$\,rad\,m$^{-2}$, with an integrated value of $121.6\pm27.5\,\upmu$Jy, is observed between the southern H{\sc ii} region D and the main body of IC\,4662.

\begin{table*}
\caption{Overview of radio continuum and polarisation properties for NGC\,3125 and IC\,4662.}
\centering
\begin{threeparttable}
\begin{tabular}{lcc}
\toprule
& NGC\,3125 & IC\,4662\\
\midrule
& Radio Continuum Properties & \\
\hline
Angular Resolution / " & $7.3 \times 3.5$  & $7.4 \times 2.9$ \\
Noise Level / $\upmu \text{Jy}$/beam & 11 & 11 \\
Path length / kpc & $2.1 \pm 0.2$ & $0.9 \pm 0.1$\\
$S_\text{tot, 1.28\,GHz} / \text{mJy}$ & $24.792 \pm 2.479$ & $28.055 \pm 2.806$\\
$S_\text{tot, 2.1\,GHz} / \text{mJy}$ & $17.306 \pm 1.731$ & $25.443 \pm 2.544$\\
$\alpha_\text{tot, 1.28-2.1\,GHz}$ & $-0.726 \pm 0.286$ & $-0.197 \pm 0.286$\\
$S_\text{nth, 1.28\,GHz} / \text{mJy}$ & $21.799 \pm 4.360$ & $17.708 \pm 3.541$\\
$S_\text{nth, 2.1\,GHz} / \text{mJy}$ & $14.247 \pm 2.850$ & $15.784 \pm 3.156$ \\
$\alpha_\text{nth, 1.28-2.1\,GHz}$ & $-0.859 \pm 0.554$ & $-0.232 \pm 0.596$\\
$\langle B_\text{eq, 2.1GHz}\rangle / \upmu \text{G}$ & $14.70^{+0.76} _{-0.69}$ &  $15.48^{+1.04} _{-0.79}$ $^\dagger$\\
\midrule
& Polarisation Properties & \\
\hline
Frequency range / MHz & [1323, 2869] & [1489, 2869] \\
n$_\text{images}$ & 1450 & 1311\\
Resolution / rad m$^{-2}$ & 83.290 & 90.206  \\
Scale / rad m$^{-2}$ & 319.437 & 319.436 \\
$\Phi$ / rad m$^{-2}$ & 106453.903 &  112381.847\\
$\lambda^2$ / rad m$^{-2}$ & 0.0217 & 0.0202 \\
PI / $\upmu$Jy  & $34.2\pm3.5$ & $121.6\pm27.5$ \\
\bottomrule
\end{tabular}
\begin{tablenotes}
\footnotesize
\item[$\dagger$] Equipartition magnetic field strength assuming a non-thermal spectral index of $-0.7$
\end{tablenotes}
\end{threeparttable}
\tablefoot{This overview includes the beam resolution, noise levels of the ATCA total radio continuum maps, and the estimated path length for both dwarf galaxies. The total and non-thermal flux densities from 1.28\,GHz (MeerKAT) and 2.1\,GHz (ATCA) data are provided, along with the calculated total and non-thermal spectral indices between these frequencies. Additionally, the equipartition magnetic field strength is reported. For the polarisation properties, we present the frequency range and number of images used for RM synthesis, the resolution in Faraday depth space, the maximum observable scale, the maximum observed Faraday depth, the observing wavelengths, and the polarised intensity. }
\label{results}
\end{table*}

\section{Discussion}
\label{discussion}

\subsection{Spectral flatness}
\label{sp_flatness}

Since we observe an extremely flat spectrum for IC\,4662 in terms of total emission and after applying the thermal correction, it is essential to investigate the underlying cause, whether it reflects inaccuracies in the thermal correction and/or both indicates dominant absorption processes.

\subsubsection{Reliability of thermal/non-thermal separation}
\label{split_trust}

The first question that arises is whether the thermal corrections developed over the years \citep[e.g.,][]{Deeg_1997,Murphy_2011} are applicable to starburst dwarf galaxies. When comparing the non-thermal spectral index maps from previous spectral index analysis studies \citep[e.g.,][]{Heesen_2008, chyzy_magnetized_2016, Basu_2017, westcott_spatially_2018}, we see that regions of star formation consistently exhibit non-thermal spectral indices close to zero. In contrast, we expect synchrotron emission from CREs that have either been freshly energized via diffusive shock acceleration \citep{Marcowith_2016} or which is subject to comparably fast synchrotron  or inverse Compton cooling \citep{Ruszkowski_2023} to show a significantly steeper spectrum.

Using the nomenclature of \citet{Heesen_2021}, the typically injected spectral index ($\gamma_{\text{inj}}$) is around $2.2$, which would result in a cooled spectral index ($\gamma_{\text{cool}}$) near $3.2$ as well, leading to an anticipated radio spectral index $(\alpha_{\nu}$) close to $-1.1$ \citep{Werhahn_2021}.
Additionally, various effects, such as diffusion losses, could further steepen the observed spectrum. If advection becomes the dominant loss mechanism, then this would lead to a radio spectral index of $\alpha_{\nu} \approx -0.6$ in the outflows, where cosmic rays are primarily advected with the gas into the halo \citep{Werhahn_2021, Heesen_2021}. However, we observe an even flatter spectrum and it is generally not expected that dwarf galaxies host such a steady wind, as discussed in Sect.~\ref{timescale}.

At lower frequencies, the spectral flux density $S(\nu)$ varies as $\nu^2$, while at higher frequencies, $S(\nu)$ follows a $\nu^{-0.1}$ dependence, which is used for the thermal correction \citep[e.g.,][]{Deeg_1997, Murphy_2011}. The frequency range where this transition occurs is known as the turnover range. When the observed galaxies, espacially IC\,4662, have H{\sc ii} regions, which are optically thick, this turnover frequency shifts to higher values, as observed in galaxies like M\,82 and NGC\,4631 \citep[e.g.,][]{Adebahr_2013, Stein_2023} for frequencies below 1\,GHz.
This shift may explain the positive spectral index observed in this dwarf galaxy. Consequently, the optical thickness of the galaxy means that H$\alpha$ corrections provide only upper limits for the radio flux calculations, with additional absorption processes occurring at longer wavelengths. 

\subsubsection{Absorption processes}
\label{absorption}

\begin{figure*}
    \centering
    \includegraphics[width=\textwidth]{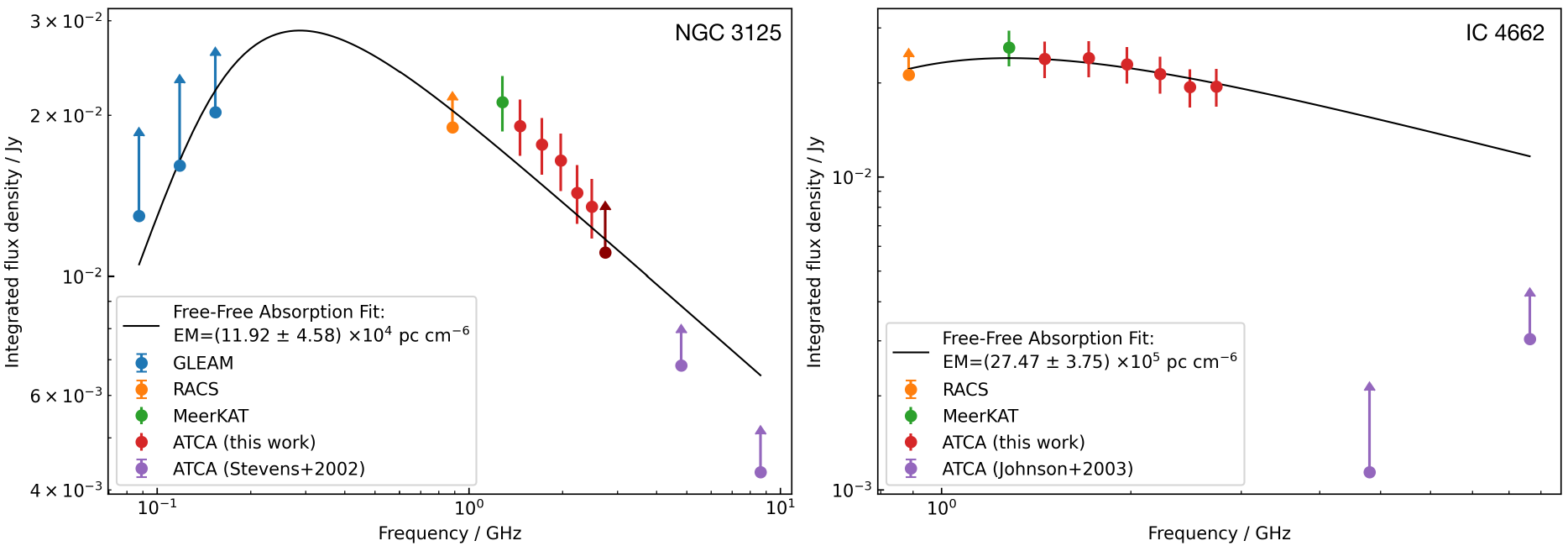}
    \caption{\textit{Left:} Integrated synchrotron emission for NGC\,3125 with the free--free absorption fit. The turnover in the spectrum can be an attribute of free--free absorption processes in the galaxy. Additionally, for NGC\,3125 there may be some flux loss at 2740\,MHz due to the use of only long baselines, therefore the data point, marked in dark red and showed as lower limit, does not contribute to the fit. 
    \textit{Right:} Integrated synchrotron emission for IC\,4662 with the free--free absorption fit. The 4.8 and 8.64\,GHz have not been taken into account for the fit. A small turnover point have been observed.}
    \label{internalthermalabsoprtion}
\end{figure*}

Synchrotron self-absorption, which flattens the radio spectrum, becomes significant at $0.15$\,MHz in the starburst dwarf galaxy IC\,4662, based on a star formation rate surface density of $\log[\Sigma_\text{SFR}/(\text{ M}_\odot \text{ yr}^{-1} \text{ kpc}^{-2})] = -1.39$ \citep{Hunter_2001}, calculated using equation~(41) from \citet{Lacki_2013}. However, due to minimal self-absorption, both total and non-thermal spectral indices remain positive, with the thermal fraction nearing $70\,\%$.
This suggests that thermal free-free absorption may be occurring, an effect also observed in other starburst dwarf galaxies such as IC\,10 and NGC\,1569 \citep{kepley_role_2010,Adebahr_2013,Basu_2017}.
This phenomenon arises in a thermal plasma, which coexists with the relativistic electrons responsible for synchrotron emission \citep[][]{Tingay_2003}. In the case of thermal free--free absorption of sources located within an ionised medium, the observed intensity can be expressed as
\begin{equation}
    S=S_0\,\biggl(\frac{\nu}{\nu_0}\biggr)^\alpha\,\,\biggl(\frac{1-e^{-\tau}}{\tau}\biggr)
\end{equation}
with the opacity $\tau$ being, 
\begin{equation}
    \tau = \frac{8.2 \times 10^{-2}\,\nu^{-2.1}\,{\rm EM}}{T_{\rm e}^{1.35}}
    \label{tau}
\end{equation}
where $\nu$ is the frequency in GHz, \textrm{EM} is the emission measure of the region in pc\,cm$^{-6}$ and $T_{\rm e}$ is the electron temperature assuming to be $10^4\,$K. The emission measure is defined as 
\begin{equation}
     \mathrm{EM}  = \int\limits_0^s n_{\rm e}^2\, {\rm d}s
    \label{em}
\end{equation}
where $n_{\rm e}$ is the  electron density\footnote{We replace the spatially varying density with its average in our calculation.} of the ionised medium in cm$^{-3}$ and s the line-of-sight through the medium in pc.

Free-free absorption is predominantly detected at lower radio frequencies, particularly within supernova remnants embedded in ionized clouds and in the spectral profiles of starburst galaxies and could be responsible for the flattening of the spectra of both dwarf galaxies. 
Furthermore, associated processes such as bremsstrahlung losses could possibly contribute to a flattening of the spectrum.

\subsection{Magnetic field structure}
\subsubsection{Role of the magnetic field}

NGC\,3125 has a compact and complex magnetic field structure with both ordered and random components. In terms of structure, the magnetic fields in dwarf galaxies like NGC\,3125 or NGC\,1569 \citep{kepley_role_2010} are often more chaotic than those in more massive galaxies \citep[e.g., edge-on galaxies,][]{Stein_2023}. This may be due to the turbulent and highly dynamic nature of the interstellar medium in dwarf galaxies, which can cause the magnetic field lines to become tangled. The gravitational potential in dwarf galaxies is weaker than in larger galaxies so that magnetic fields can play a stronger role in shaping the gas dynamics of these galaxies \citep[e.g.,][]{chyzy_regular_2000, kepley_role_2010, chyzy_magnetized_2016}. 

Magnetic field amplification in NGC\,3125 could arise from processes tied to star formation and galactic evolution. High rates of supernova explosions and intense star formation generate turbulence and gas motions, which stretch and twist magnetic field lines, amplifying them. This amplification impacts the interstellar medium, influencing gas dynamics and potentially increasing synchrotron radiation and CR production \citep{Siejkowski_2010}. 

Several mechanisms have been proposed for this magnetic field generation. The supernova-driven dynamo suggests that supernovae-induced turbulence and rotational flows in the ISM amplify the magnetic field, a concept supported by observations and simulations \citep{Gressel_2008,Gent_2023}. Alternatively, the gravitational collapse of the proto-galaxy drives turbulence, which amplifies the magnetic field through a small-scale dynamo \citep{Pakmor_2017,Pakmor_2024,Pfrommer_2022,Liu_2022}. The convective wind-driven hypothesis proposes that convective winds generate magnetohydrodynamic turbulence, leading to similar amplification \citep{chyzy_magnetized_2016}, which we showed is not the dominant process in NGC\,3125 and IC\,4662. The magnetic field orientation outside of IC\,4662 extends from north-west to south-east, suggesting possible inflow or outflow dynamics.

In both dwarf galaxies, traditional large scale $\alpha-\Omega$ dynamo mechanisms are unlikely due to chaotic gas motions and slow rotation. Instead, NGC\,3125 potentially hosts magnetic fields shaped by small-scale fluctuating dynamo processes.

\subsubsection{Interaction between the H{\sc ii} region and the main body in IC\,4662}
\label{interaction}

There is a controversial discussion whether the southern H{\sc ii} region D is physically detached from the main body or is still interacting with it \citep[e.g.,][]{Eymeren2010, Hidalgo_2001}.
\citet{Eymeren2010} suggest that both region D and the main body reside within the same H{\sc i} disc, indicating a direct physical connection. On the other hand, \citet{Hidalgo_2001} argues that the differing chemical abundances between region D and the main body suggest that region D may not be part of IC 4662. 

With this polarimetry study of IC\,4662, observations reveal a magnetic field extending between region D and the main body, indicating an ongoing physical interaction. This magnetic connection strengthens the case that region D is not entirely detached but remains dynamically linked to the main body, despite the observed differences in chemical composition.

\subsection{Depolarisation processes}
\label{depolarisation}

In NGC\,3125, the observed regions exhibiting flat spectra (see Fig.~\ref{si_sc_ngc}) and no corresponding polarised emission (see Fig.~\ref{PI_ngc}) can be attributed to beam depolarisation of an already small non-thermal contribution to the observed radio emission. 

In Fig.~\ref{PI_ic}, the absence of polarised emission in the main body of IC\,4662 can be explained by several depolarisation mechanisms. Turbulent magnetic fields within IC\,4662 may play a significant role, as disordered or small-scale turbulence in the magnetic field could produce a mix of polarisation angles within each beam element, leading to depolarisation.
Another possibility is Faraday depolarisation, where polarised synchrotron radiation passes through magnetised ionised gas along the line of sight. Variations in electron density and magnetic field strength cause differing amounts of Faraday rotation across the beam, resulting in partial or complete cancellation of polarisation vectors, thus reducing the detected polarisation. Another contributing factor could be Faraday depth depolarisation, where different Faraday rotation measures along the line of sight cause destructive interference of polarised signals, further diminishing the observed emission.

\section{Conclusions} 
\label{conclusion}

We present the first radio continuum polarimetry study to date of two of the brightest starburst dwarf galaxies, NGC\,3125 and IC\,4662, in the IRAS RGBS sample. The combination of 16\,cm observations with ATCA and already existing archive data from MeerKAT at 1.28\,GHz \citep{Condon_2021} provides the opportunity to gain insights in starburst galaxies and leads us to the following conclusions.

\begin{enumerate}
    \item We detect radio continuum emission from NGC\,3125 and IC\,4662 at 1.28\,GHz and 2.1\,GHz. For NGC\,3125, we discover an extension of the synchrotron halo perpendicular to the disk, resulting in a symmetric envelope, at both frequencies. IC\,4662 has a complex structure consisting of two central star-forming regions surrounded by a diffuse emission halo with overall predominantly thermal emission.

    \item The box integration method demonstrates the profile of the total and non-thermal spectral index, exhibiting no significant difference between the two. This raises questions about the reliability of the thermal correction, particularly in relation to the consideration of an optically thin medium for dwarfs or an optically thick medium, which would result in a different correction. 
    
    \item The overall integrated non-thermal spectral index is $-0.63 \pm 0.03$ for NGC\,3125 and $-0.5 \pm 0.07$ for IC\,4662, respectively. These galaxies exhibit comparable behaviour to that observed in other starburst dwarf galaxies, such as NGC\,1569 \citep[$\alpha_\text{nth}=-0.77 \pm 0.03$,][]{kepley_role_2010} and IC\,10 \citep[$\alpha_\text{nth}=-0.55 \pm 0.04$,][]{Basu_2017}. 
    IC\,4662 shows overall a positive total spectral index, which could be associated to free--free absorption.
    
    \item We detect weak polarised emission at 2.1\,GHz for NGC\,3125, resulting in a degree of polarisation ranging from 0.1\,\% to 1.8\,\%. This is comparable to the findings reported for IC\,10 \citep[with 1.4\,\%,][]{chyzy_regular_2000}. 
    In the case of IC\,4662, no polarised emission within the galaxy was detected, which could be the result of depolarisation in the main body of the galaxy. However, polarised emission was observed between the southern H{\sc ii} region and the main body of the dwarf galaxy, indicating potential interaction between the both which is illuminated by a magnetized radio emitting plasma.
    
    \item The CRE cooling timescale $\tau_\text{cool}$ is much smaller than $\tau_\text{esc}$ in both dwarf galaxies, implying a steepened CRE spectrum, which would be steepened by one in steady state \citep{Ruszkowski_2023}. Explaining their flat spectra requires additional processes such as free-free emission at high frequencies and absorption at low frequencies, in line with expected starburst galaxy behaviour \citep[e.g., in M\,82, ][]{Adebahr_2013, Werhahn_2021}.    
    
\end{enumerate}

Using ATCA and MeerKAT to study magnetised galactic outflows in the dwarf starburst galaxies NGC\,3125 and IC\,4662 improves our understanding of the interstellar medium and non-thermal processes in these starburst environments, and possible will allow us to draw conclusions on CR feedback mechanisms.

\begin{acknowledgements}
ST, BA, DJB, MS and RJD acknowledge the support from the DFG via the Collaborative Research Center SFB1491 \textit{Cosmic Interacting Matters - From Source to Signal}. CP acknowledges support by the European Research Council under ERC-AdG grant PICOGAL-101019746. 
PK and RJD acknowledge the support of the BMBF project 05A23PC1 for D-MeerKAT.
The Australia Telescope Compact Array is part of the Australia Telescope National Facility (https://ror.org/05qajvd42) which is funded by the Australian Government for operation as a National Facility managed by CSIRO. We acknowledge the Gomeroi people as the Traditional Owners of the Observatory site.
The MeerKAT telescope is operated by the South African Radio Astronomy Observatory, which is a facility of the National Research Foundation, an agency of the Department of Science and Innovation.
\end{acknowledgements}

\bibliography{literatur}
\bibliographystyle{aa}
\appendix
\section{Non-thermal emission}
\label{NT}
The non-thermal maps for 1.28\,GHz and ATCA 2.1\,GHz of each galaxy are shown in this section. The maps are overlaid with $3\sigma$ contours of the non-thermal emission.

\begin{figure*}[h!]
    \centering
    \includegraphics[width=\linewidth]{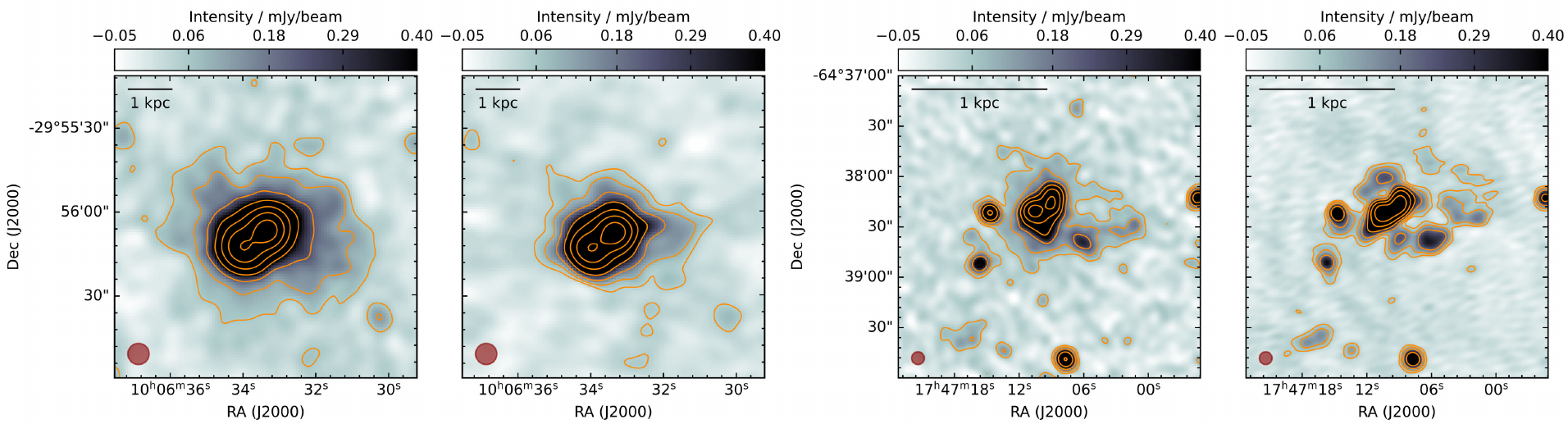}
    \caption{\textit{Left:} Non-thermal emission of NGC\,3125 at 1.28\,GHz with overlaid contours starting at $3\sigma$ and increase by factor of 2 ($\sigma = 20\,\upmu$Jy/beam). \textit{Middle left:} Non-thermal emission of NGC\,3125 at 2.1\,GHz with overlaid contours starting at $3\sigma$ and increase by factor of 2 ($\sigma = 15\,\upmu$Jy/beam).
    \textit{Middle Right:} Non-thermal emission of IC\,4662 at 1.28\,GHz with overlaid contours starting at $3\sigma$ and increase by factor of 2 ($\sigma = 20\,\upmu$Jy/beam). \textit{Right} Non-thermal emission of IC\,4662 at 2.1\,GHz with overlaid contours starting at $3\sigma$ and increase by factor of 2 ($\sigma = 15\,\upmu$Jy/beam).
    The $7.6"$ circular beam appear in the lower left corner.}
    \label{syn}
\end{figure*}

\section{Thermal fraction}
\label{tf_sect}
The thermal fraction maps for 1.28\,GHz and ATCA 2.1\,GHz of each galaxy are shown in this section. The maps are overlaid with $3\sigma$ contours of the non-thermal emission.

\begin{figure*}[h!]
    \centering
    \includegraphics[width=\textwidth]{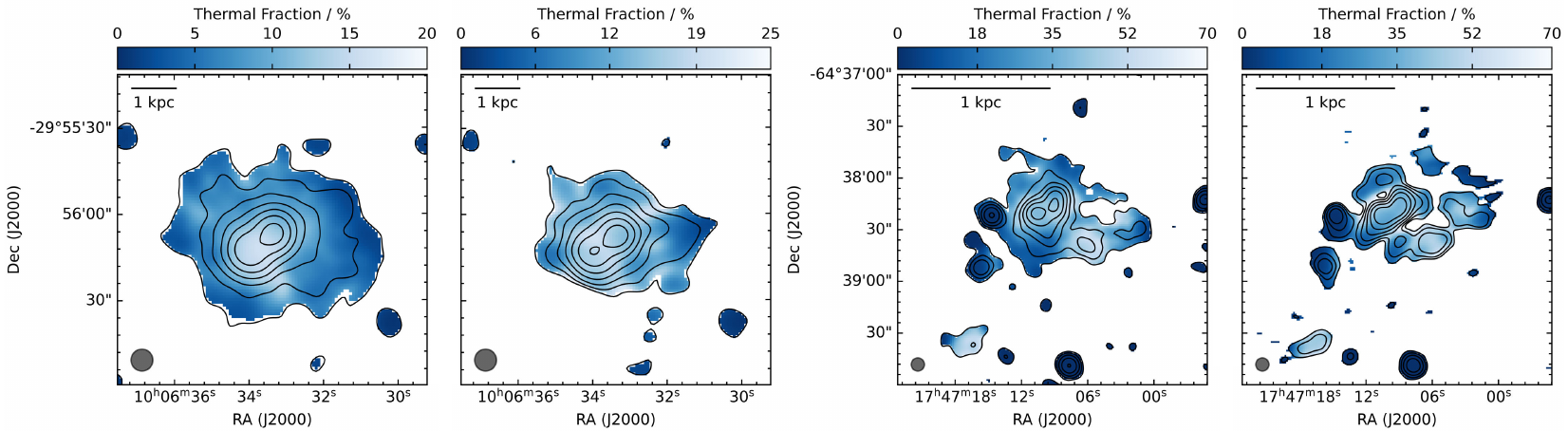}
    \caption{\textit{Left:} Thermal fraction of NGC\,3125 at 1.28\,GHz with overlaid contours of the non-thermal emission starting at $3\sigma$ and increase by factor of 2 ($\sigma= 20\,\upmu$Jy/beam). 
    \textit{Middle left:} Thermal fraction of NGC\,3125 at 2.1\,GHz with overlaid contours of the non-thermal emission starting at $3\sigma$ and increase by factor of 2 ($\sigma= 15\,\upmu$Jy/beam).
    \textit{Middle right:} Thermal fraction of IC\,4662 at 1.28\,GHz with overlaid contours of the non-thermal emission starting at $3\sigma$ and increase by factor of 2 ($\sigma= 20\,\upmu$Jy/beam). 
    \textit{Right:} Thermal fraction of IC\,4662 at 2.1\,GHz with overlaid contours of the non-thermal emission starting at $3\sigma$ and increase by factor of 2 ($\sigma= 15\,\upmu$Jy/beam).
    The 7.6" circular beam appear in the lower left corner.}
    \label{tf}
\end{figure*}

\section{Non-thermal spectral index}
\label{spix_nth}
The non-thermal spectral index maps of each galaxy are shown in this section. The maps are overlaid with $5\sigma$ contours of the non-thermal emission.

\begin{figure*}[h!]
    \centering
    \includegraphics[width=\linewidth]{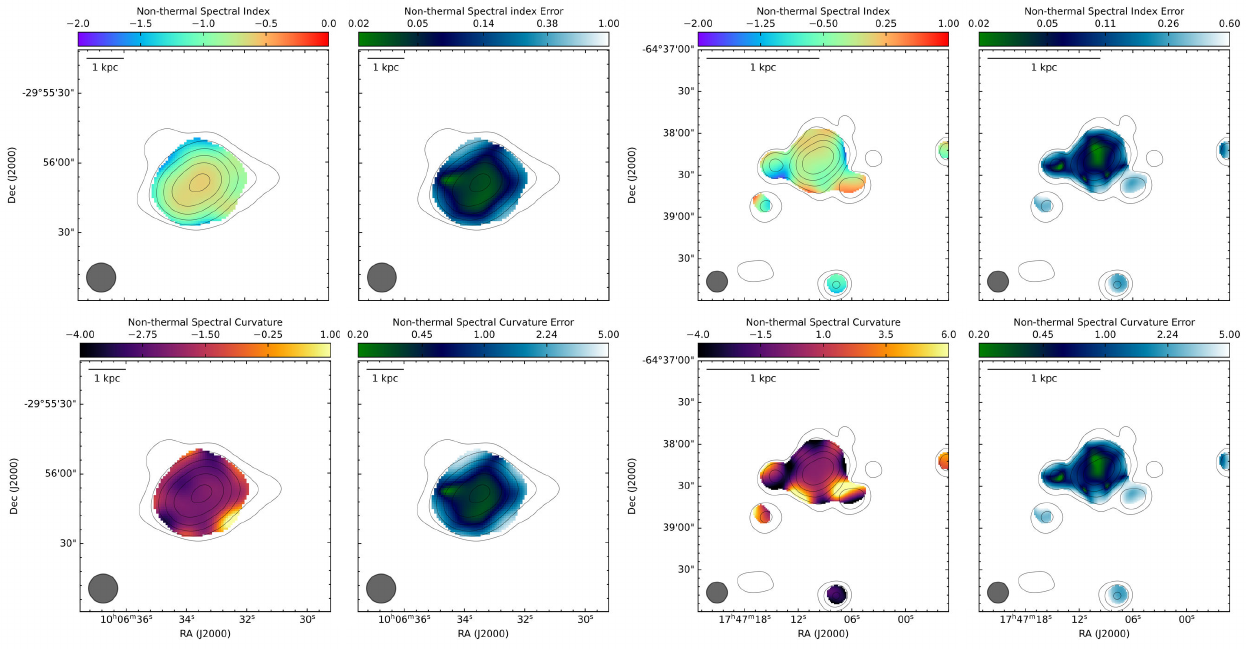}
    \caption{ 
    \textit{Top:} Fitted in-band non-thermal spectral index and its error of NGC\,3125 (left) and IC\,4662 (right) between the frequency ranges 1460\,MHz and 2740\,MHz with overlaid non-thermal emission at 2.1\,GHz with $\sigma =20\,\upmu$Jy/beam for NGC\,3125 (left) and $\sigma =23\,\upmu$Jy/beam for IC\,4662 (right).
    \textit{Bottom:} Fitted in-band non-thermal spectral curvature and its error of NGC\,3125 (left) and IC\,4662 (right) between the frequency ranges 1460\,MHz and 2740\,MHz with overlaid non-thermal emission at 2.1\,GHz with $\sigma =20\,\upmu$Jy/beam for NGC\,3125 (left) and $\sigma =23\,\upmu$Jy/beam for IC\,4662 (right). 
    The contours increasing by a factor of 2 starting at $5\sigma$. The circular beam of $12.5"$ for NGC\,3125 (left) and $15"$ for IC\,4662 (right) is shown in the left corner.}
    \label{si_nt}
\end{figure*}

\section{Equipartition magnetic field strength}
\label{MF}
A conventional method of calculating the magnetic field strength involves assuming equipartition between the energy densities of CRs and the magnetic field even though there are large uncertainties associated with these assumptions \citep{Pfrommer_2004MNRAS, kepley_role_2010, Seta_2019, Ruszkowski_2023, Dacunha_2024}. Assuming that the thermal/non-thermal split is valid and that the equipartition assumptions can be applied to starburst dwarf galaxies, the equipartition magnetic field strength can be calculated following \citet{Beck_2005} as 
\begin{equation}
    B_\text{eq}=\Biggl(\frac{4\pi (2\alpha-1)(K_0+1)I_\nu E_\text{p}^{1+2\alpha}(\frac{\nu}{2c_1})^{-\alpha}}{(2\alpha +1)c_2(\alpha)lc_4(i)}\Biggr)^\frac{1}{(3 - \alpha)}
    \label{equipartition_eq}
\end{equation}
with $\alpha$ as the non-thermal spectral index and $I_\nu$ as the corresponding synchrotron intensity at a given frequency $\nu$.
We adopted a proton-to-electron ratio of energy densities of $K_0 = 100$ for each galaxy, consistent with the values used in previous studies \citep[e.g.,][]{chyzy_magnetized_2016,Basu_2017,heesen_nearby_2022,Stein_2023}.
The inclination $i$ and the path length $l$ are taken from Tables~\ref{basics} and \ref{results}. A detailed description of $c_1,c_2$ and $c_4$ can be taken from \citet{Beck_2005}. 
The equipartition formula for simple power law energy spectra becomes singular at $\alpha=-0.5$ and the magnetic field strength cannot be determined in these cases. 
The path length $l$ is difficult to determine as the geometry of galaxies is not well known. One possible method to estimate it is to assume a spherical symmetric structure and to use the extent of H$\alpha$ and radio emission to calculate their mean as the estimation of $l$, as applied in this study (Table~\ref{results}).

Using the Bayesian approach \citep{Zychowicz2025} we also calculated the mean equipartition magnetic field strength based on the mean synchrotron intensity of galaxies. In this method, magnetic field is a random variable treated as the posterior distribution, which automatically provides uncertainties in the estimated magnetic field strength. We calculated the mean nonthermal emission at 2.1\,GHz within the area delineated by the emission above the 5 rms map level. We used uncertainties of 10\% for the parameters $I_{\nu}$, $l$, $K_0$, assuming their Gaussian distribution. Under the assumption of a purely random magnetic field, we obtained the magnetic field strength for NGC\,3125 of $14.70^{+0.76}_{-0.69}\,\upmu$G (Table~\ref{results}). If we instead assumed that the field is entirely ordered, the estimated magnetic field increases to $17.45^{+0.92}_{-0.83}\,\upmu$G. Using the conventional equipartition equation (Eq.~\eqref{equipartition_eq}) and the BFIELD program from \citet{Beck_2005}, we obtain a mean field strength of $16.3\,\upmu$G. These results are in close agreement, indicating consistency across the different methods.
They are also similar to those found in other starburst dwarf galaxies, such as NGC\,1569, NGC\,4449, IC\,10, and NGC\,253 \citep{kepley_role_2010, chyzy_regular_2000, chyzy_magnetized_2016, Basu_2017, Heesen_2008}. 

However, several potential sources of errors should be considered. One is related to uncertain knowledge of galaxy geometry and the determination of $l$. If the path length is doubled, the resulting equipartition magnetic field strength differs by about $17\%$ from the original result. Another important assumption in this analysis is the use of $K_0=100$, as adopted by \citet{Heesen_2023}. There may be discrepancies in this value when applied to dwarf galaxies because of the potential for a more turbulent magnetic field, which may be less able to stabilize effectively \citep{hindson_radio_2018}. Additionally, the spectral index is a critical input parameter. Values approaching $-0.54$ can significantly increase the calculated equipartition magnetic field strength. This is relevant for our analysis, as we observe a flat spectrum at 2.1\,GHz, possibly due to thermal free-free absorption.

\section{Distribution of Faraday rotation measure}
\label{frm}
\begin{figure*}[h!]
    \centering
    \includegraphics[width=0.75\linewidth]{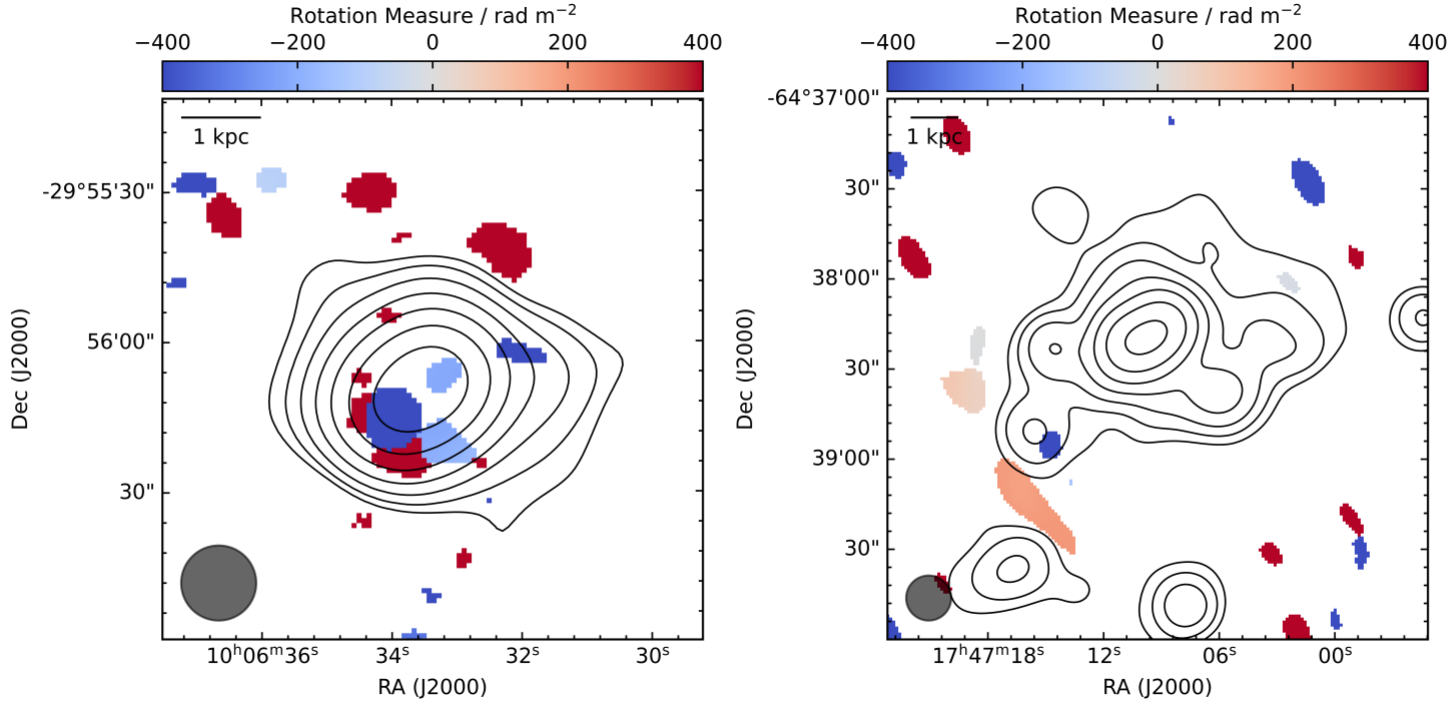}
    \caption{\textit{Left:} Distribution of RM of NGC\,3125 at 2.1\,GHz, already corrected for the foreground ($\sigma = 22\,\upmu$Jy/beam). \textit{Right:} Distribution of RM of IC\,4662 at 2.1\,GHz, already corrected for the foreground ($\sigma = 40\,\upmu$Jy/beam).  The contours represent the non-thermal emission of each galaxy at 2.1\,GHz, convolved to circular beam of $15"$ with level of (3, 6, 12, 24, 48, 96) $\times \sigma$. The circular beam of $15"$ is shown in the left corner.}
    \label{RM}
\end{figure*}

The left panel of Fig.~\ref{RM} shows the distribution of the RM of NGC\,3125 at 2.1\,GHz, corrected for the RM foreground of $-45.8\,\text{rad m}^{-2}$ \citep{Oppermann_2012}. The most notable RM structure in the galaxy is situated in the central regions, extending towards the south. This feature, characterised by a flat spectrum, represents the most concentrated RM region within the galaxy.
In this south area, we obtain a RM value of approximately $-434 \pm 3.7\,\text{rad m}^{-2}$, while the north-east region has a RM of approximately $-168 \pm 3.6 \,\text{rad m}^{-2}$. Large RM values outside these regions, where the RM distribution is much more patchy, is not reliable that these region correspond to filament of the galaxy. 
The RM islands appear patchy at beam scale, even within the main body. The sharp $-400$ to $+400\,\text{rad m}^{-2}$ jump raises uncertainty, particularly given potential complexity in the FD spectra. Large Faraday depolarization could indicate multiple RM components, complicating peak identification. 

In the right panel of Fig.~\ref{RM}, the RM distribution of IC\,4662 at 2.1\,GHz is shown, corrected for the RM foreground of $49.3\,\text{rad m}^{-2}$ \citep{Oppermann_2012}. If we count these 5$\sigma$ clipped patches as a detection, we see in region C a RM of approximately $-21 \pm 3.9\,\text{rad m}^{-2}$. In the south of region E, we detect a RM of approximately $-231 \pm 3.7\,\text{rad m}^{-2}$. Between region E in the main body of IC\,4662 and region D to the south, we observe an RM of $190 \pm 5.3,\text{rad m}^{-2}$, suggesting a potential magnetic field interaction between these two regions.

Following \citet{Burn_1966}, the magnitude of the Faraday effect is described by $\chi -\chi_0 =\text{RM}\,\lambda^2$, with
\begin{equation}
    \text{RM}\approx 0.81 \frac{\mathrm{rad}}{\mathrm{m}^2}\int\limits^{\text{observer}}_{\text{source}} n_{\rm e} \,B_\parallel \, \text{d}l
    \label{rm}
\end{equation}

Here $\chi$ and $\chi_0$ in rad are the observed and emitted position angles, respectively, of a linearly polarised radio wave, $\lambda$ in m$^2$ the observing wavelength, $n_{\rm e}$ in cm$^{-3}$ the free electron density, $B_\parallel$ in $\upmu$G is the magnitude of the magnetic field vector projected along the line-of-sight, and d$l$ in pc is an infinitesimal distance interval along the line-of-sight.

To derive an accurate magnetic field strength along the line-of-sight, we generally need to know the clumping factor, as EM is highly sensitive to electron clumping, while RM is not \citep{Hutschenreuter_2023}.
But RM derived from RM synthesis from an emitting and Faraday rotating medium mostly comes from regions along the line-of-sight that contribute the largest polarized emission \citep{Basu_2019}. This introduces additional complexity in estimating the filling factor $f_v$, which is equivalent to the dispersion measure, particularly in external galaxies. This can be circumvented if there is some information on Faraday depolarization, i.e., $\sigma_{\rm RM}$, and then quote the field strengths in terms of the filling factor $f_v$ \citep{Basu_2017_rm}. 

However, for a preliminary estimate of the magnetic field, we proceed without incorporating the filling factor $f_v$. Using Eq.~\eqref{rm}, with the information of the electron density $n_{\rm e}$ through the emission measure EM, described in the Sect.~\ref{absorption}, which has been taken from Table \ref{timescales}, we can calculate the magnetic field along the line-of-sight. For NGC\,3125, we have calculated an integrated magnetic field strength $B_\parallel = 7.15\,\upmu$G. In case of IC\,4662, we calculate an integrated magnetic field strength $B_\parallel = 2.2\,\upmu$G between the southern H{\sc ii} region and the main body.

\section{Cosmic ray electron losses}
\label{timescale}
In galaxies exhibiting a very flat spectrum, Coulomb interactions and bremsstrahlung losses could become particularly significant, especially in dense environments \citep[e.g. M\,82 and M\,51,][]{Adebahr_2013, Gajovic_2024}. While Coulomb losses are primarily effective at lower CRE energies, they are largely irrelevant at the higher energies associated with synchrotron-emitting CREs \citep{Petrosian_2001, Ruszkowski_2023}. 
The non-thermal bremsstrahlung lifetime $\tau_\text{brems}$ can be expressed as follows \citep{Werhahn_2021}:
\begin{equation}
    \tau_\text{brems} = \frac{E_{\rm e}}{4\alpha r_0^2 n_{\rm e} \beta_{\rm e} \gamma_{\rm e} m_{\rm e} c^3} \biggl[ \log(2\gamma_{\rm e})-\frac{1}{3}\biggr]^{-1},
\end{equation}
where the normalised electron velocity is denoted by $\beta_{\rm e}=\varv_{\rm e}/c$, $r_0$ is the electron radius, $\alpha$ is the fine-structure constant, $c$ is the speed of light, and $m_{\rm{ e}}$ is the electron mass. For highly relativistic electrons, we can assume a Lorentz factor $\gamma_{\rm e}$ of $10^4$ \citep{Ruszkowski_2023, Werhahn_2021}, therefore electrons that emit synchrotron emission at GHz frequencies in $\upmu$G magnetic fields have a typical energy of $E_{\rm e} = \gamma_{\rm e} m_{\rm e} c^2 \approx 5$\,GeV. 

To determine the average thermal electron density $n_{\rm e}$, we fit the free-free absorption function (see Sect.~\ref{absorption}), where we assume that the line-of-sight length equals the path length (from Table~\ref{results}). We used for both galaxy, the H$\alpha$ corrected non-thermal spectral indices of $-0.63$ for NGC\,3125 and $-0.5$ for IC\,4662 based on their non-thermal emissions, as shown in the integrated spectral energy distribution (Fig.~\ref{sed_int}). The fitted emission measure EM and the calculated the average thermal electron density $n_{\rm e}$ derived from Eq.~\eqref{em}, are provided in Table \ref{timescales}.

The timescales of synchrotron losses $\tau_{\text{syn}}$ and inverse Compton losses $\tau_{\text{IC}}$ can be defined as in \citet{Ruszkowski_2023}:
\begin{equation}  
    \tau_{\text{syn}} = \frac{6 \pi m_{\rm e} c}{\sigma_{\rm T} B^2 \gamma_{\rm e}}
\end{equation} 
and 
\begin{equation}
    \tau_{\rm IC} = \frac{6 \pi m_{\rm e} c}{\sigma_{\rm T} B_{\rm ph}^2 \gamma_{\rm e}}
\end{equation}

where the photon energy density $\epsilon_{\rm ph}$, can be expressed with the equivalent magnetic field strength, $B_{\rm ph} = \sqrt{8\pi \epsilon_{\rm ph}}$. 

The photon energy density is the sum of the CMB and starlight, $\epsilon_{\rm ph}=\epsilon_{\rm CMB}+\epsilon_{\rm *}$, with $\epsilon_{\rm CMB} \approx [3.2(1+z)^2]^2/(8\pi)$ \citep{Ruszkowski_2023}.
For the stellar radiation term, $\epsilon_{\rm *}$, we follow the approach of \citet{Werhahn_2021}, assuming that the UV light emitted by young stellar populations is re-emitted in the FIR. We use the relation from \citet{Kennicutt_1998}, 
\begin{equation}
    \frac{L_{\rm FIR}}{L_\odot} = 7.5 \times 10^9 \frac{\rm SFR}{M_\odot\,{\rm yr^{-1}}}
\end{equation}
to express $\epsilon_{\rm *}$ as  
\begin{equation}
\epsilon_{\rm *} = \frac{L_{\rm FIR}}{4\pi R^2 c}
\end{equation}
where $R$ is radius of the galaxy, which is in our model half of the path length and SFR is the star formation rate to the galaxy.
The magnetic field strength  $B$, calculated under the equipartition assumption (see Appendix~\ref{MF}), is listed in Table~\ref{results}. For IC\,4662, only an upper limit for $B$ can be calculated, constrained by the spectral index limit. This suggests that thermal electrons likely contribute to the observed spectral flattening. As shown in Table~\ref{timescales}, synchrotron and IC losses dominate the CRE cooling in NGC\,3125, while bremsstrahlung, synchrotron and IC losses are all of similar order in IC\,4662.

It is now possible to determine whether CREs escape through a galactic wind as a result of advection or if they cool so fast, that they cannot escape the galaxy. By comparing the relative importance of these two processes, it is possible to determine the dominant mechanism. The cooling timescale for a continuous injection of CRs can be calculated by combining all the timescales:
\begin{equation}
    \tau_\text{cool}^{-1} =  \tau_\text{syn}^{-1} + \tau_\text{IC}^{-1} + \tau_\text{brems}^{-1}.
\end{equation}
Assuming that CR streaming at the Alfvén speed dominates the spatial transport of CRs, the escape timescale due to galactic wind can be written as follows,
\begin{equation}
    \tau_\text{esc} \sim \frac{h}{\varv_{\rm A}}.
\end{equation}
where $h$ denotes the disc scale height, representing half of the path length in our model of a spherically symmetric galaxy, while $\varv_{\rm A}=B/\sqrt{4\pi \rho}$ indicates the Alfvén speed.

As shown in Table~\ref{timescales}, the CRE cooling time scale $\tau_\text{cool}$ is significantly smaller than $\tau_\text{esc}$ for both dwarf galaxies. Assuming that the CRE injection balances their cooling losses, we conclude that the CRE population is in steady state and that their injection spectral index is steepened by one \citep{Ruszkowski_2023}. Hence, we need to consider other processes such as free-free emission at high frequencies and absorption at low frequencies to explain their flat spectral indices \citep{Werhahn_2021}. This is also in accordance with the expected behaviour for starburst galaxies \citep[e.g. M82,][]{Adebahr_2013}.

It is important to note that H$\alpha$ filaments are observed to extend farther into the halo compared to the detected radio emission (Fig.~\ref{PI_ngc}). This discrepancy could be attributed to rapid cooling of CRe at the outskirts of the galaxy, or it might reflect limitations in the sensitivity of the radio observations.

\begin{table}
\centering
\caption{Overview of the timescales of the CREs, the fitted emission measure, the calculated electron density and Alfv\'en speed.}
\begin{tabular}{lcc}
\toprule
 & NGC\,3125 & IC\,4662 \\
 \midrule
 EM / ($10^4$ pc\,cm$^{-3}$) & 11.9 &  275 \\
$\langle n_{\rm e} \rangle$ / cm$^{-3}$ & 7.2 & 55.2 \\
$\varv_{\rm A}$ / (km/s) & 11.9 & 4.5 \\
 \hline
$\tau_\text{syn}$ / Myrs & 1.5 & 1.4 \\
$\tau_\text{IC}$ / Myrs & 0.3 & 0.4 \\
$\tau_\text{brems}$ / Myrs & 6.5 & 0.8 \\
\hline
$\tau_\text{cool}$ / Myrs & 0.2 & 0.2\\
$\tau_\text{esc}$ / Myrs & 96.6 & 99.1 \\
\bottomrule
\end{tabular}
\tablefoot{All values were calculated for the 2.1\,GHz observations.}
\label{timescales}
\end{table}

\label{lastpage}
\end{document}